\documentclass[twocolumn,twocolappendix]{aastex701}

\usepackage{float,graphicx,amsmath,multirow}
\usepackage[version=4]{mhchem}
\usepackage{booktabs}
\usepackage{enumitem}
\usepackage{wrapfig}
\usepackage{threeparttable}
\usepackage{comment}
\usepackage{upgreek}
\usepackage{wasysym}

\begin{document}

\title{Detection of the Polycyclic Aromatic Hydrocarbon Phenalene (\ce{C13H10}) in the Very Low Luminosity Object (VeLLO) MC27/L1521F}


\author[0000-0002-0332-2641]{Gabi Wenzel}
\affiliation{Department of Chemistry, Massachusetts Institute of Technology, Cambridge, MA 02139, USA}
\email{gwenzel@mit.edu}
\affiliation{Center for Astrophysics \textbar{} Harvard \& Smithsonian, Cambridge, MA 02138, USA}
\email{gabi.wenzel@cfa.harvard.edu}
\author[0000-0001-8134-5681]{Thomas H. Speak}
\affiliation{Department of Chemistry, University of British Columbia, Vancouver, BC V6T 1Z1, Canada}
\email{}
\author[0000-0003-2760-2119]{Ci Xue}
\affiliation{National Radio Astronomy Observatory, Charlottesville, VA 22903, USA}
\affiliation{NSF–Simons AI Institute for Cosmic Origins, Austin, TX, 78712, USA}
\email{}


\author[0000-0003-4179-6394]{Edwin A. Bergin}
\affiliation{Department of Astronomy, University of Michigan, Ann Arbor, MI 48109, USA}
\email{}
\author[0000-0003-0799-0927]{Andrew M. Burkhardt}
\affiliation{Department of Earth, Environment, and Physics, Worcester State University, Worcester, MA 01602, USA}
\email{}
\author[0000-0001-8233-2436]{Martin A. Cordiner} 
\affiliation{Astrochemistry Laboratory and the Goddard Center of Astrobiology, Solar System Exploration Division, NASA Goddard Space Flight Center, Greenbelt, MD 20771, USA }
\email{}
\author[0009-0007-2048-2907]{Miya Duffy}
\affiliation{Department of Chemistry, Massachusetts Institute of Technology, Cambridge, MA 02139, USA}
\email{}
\author[0000-0001-5020-5774]{Zachary T. P. Fried}
\affiliation{Department of Chemistry, Massachusetts Institute of Technology, Cambridge, MA 02139, USA}
\email{}
\author[0000-0002-6667-7773]{Andrew Lipnicky}
\affiliation{National Radio Astronomy Observatory, Charlottesville, VA 22903, USA}
\email{}
\author[0000-0002-5171-7568]{Christopher N. Shingledecker}
\affiliation{Department of Chemistry, Virginia Military Institute, Lexington, VA 24450, USA}
\email{}

\author[0009-0002-6372-9926]{Reace H. J. Willis}
\affiliation{Department of Chemistry, University of British Columbia, Vancouver, BC V6T 1Z1, Canada}
\email{}


\author[0000-0001-9479-9287]{Anthony J. Remijan}
\affiliation{National Radio Astronomy Observatory, Charlottesville, VA 22903, USA}
\email{}
\author[0000-0001-9142-0008]{Michael C. McCarthy}
\affiliation{Center for Astrophysics \textbar{} Harvard \& Smithsonian, Cambridge, MA 02138, USA}
\email{}
\author[0000-0003-1254-4817]{Brett A. McGuire}
\affiliation{Department of Chemistry, Massachusetts Institute of Technology, Cambridge, MA 02139, USA}
\affiliation{National Radio Astronomy Observatory, Charlottesville, VA 22903, USA}
\email{}
\author[0000-0002-0850-7426]{Ilsa R. Cooke}
\affiliation{Department of Chemistry, University of British Columbia, Vancouver, BC V6T 1Z1, Canada}
\email{}


\correspondingauthor{G. Wenzel, B. A. McGuire, I. R. Cooke}
\email{gwenzel@mit.edu, brettmc@mit.edu, icooke@chem.ubc.ca}

\begin{abstract}
To date, 14 polycyclic aromatic hydrocarbons (PAHs) ranging in size from two to seven (including five- and six-membered) carbon rings have been detected in the starless dense core TMC-1 CP within the Taurus molecular cloud. Their detection raises questions about the distribution of PAHs in the cold interstellar medium (ISM) and their evolution during star formation. Here, we present the first interstellar detection of a three-ring PAH outside of TMC-1 CP. We detect phenalene (\ce{C13H10}), a compact, peri-fused PAH, in the dense core MC27/L1521F, a molecular cloud in Taurus containing a very low-luminosity object (VeLLO). We compare the abundances of phenalene in the two sources with respect to the single-ring aromatic benzonitrile, and find that it is enhanced by a factor of four in MC27/L1521F. We discuss the implications for possible formation and destruction pathways in the two sources. These findings further support the widespread abundance of PAHs throughout the cold ISM and are consistent with survival, inheritance, or replenishment during the earliest stages of star formation.
\end{abstract}


\section{Introduction} 

Polycyclic aromatic hydrocarbons (PAHs) are key components of the interstellar medium (ISM) and have been established as major carbon reservoirs. While their interstellar presence has been assumed since the 1980s \citep{leger1984,allamandola1985}, unambiguous evidence was delivered recently with the detections of cyano-functionalized PAHs (CN-PAHs) in the dark and cold Taurus molecular cloud (TMC-1) ranging from two to up to seven perifused aromatic rings, namely naphthalene, acenaphthylene, pyrene, and coronene \citep{mcguire2021,cernicharo2024,cernicharo2026,wenzel2024,wenzel2025,wenzel2025c}. These results stem from two sensitive line surveys using single dish radio telescopes; the Green Bank Telescope (GBT) Observations of TMC-1: Hunting for Aromatic Molecules (GOTHAM;~\citealt{mcguire2020}) and Q-band Ultrasensitive Inspection Journey to the Obscure TMC-1 Environment (QUIJOTE;~\citealt{cernicharo2021b}) projects.

Depending on their symmetry and substitution pattern, bare or unsubstituted PAHs possess small to zero permanent electric dipole moments, making the CN-PAHs attractive proxies with bright rotational transitions \citep{mcguire2018a,cooke2020,mccarthy2026a}. Nevertheless, the bare PAHs 1H-indene, 1H-cyclopent[cd]indene, and 1H-phenalene, with permanent electric dipole moments of approximately 0.5--0.8\,D, have been detected in TMC-1 CP (Cyanopolyyne Peak) thanks to their radio emission and large abundance  \citep{burkhardt2021,cernicharo2021,fuentetaja2026,cabezas2025}. The substantial extent of the carbon reservoir held by PAHs in TMC-1 CP was rather unexpected, as was the flat abundance distribution irrespective of size or molecular complexity \citep{wenzel2025c, rivilla2026, agundez2026}, defying the near universal trend of decreasing abundance with increasing size within a molecular ``family" \citep{loomis2016, loomis2021}.

TMC-1 CP is a starless core before protostar formation, but not the only source hosting aromatic molecules. The single-ring aromatic molecule benzonitrile has been identified toward sources located in Taurus, Serpens, Lupus, and Aquila \citep{burkhardt2021a,agundez2023} and most recently in the Central Molecular Zone \citep{rivilla2026}. Additionally, the two-ring molecule 1H-indene was detected toward the quiescent starless core L1495B \citep{agundez2023}, a source in Taurus exhibiting increased abundances of oxygen-bearing sulfur species when compared to other starless or prestellar cores \citep{scholler2026}. A representative object transitioning from the prestellar core stage to the protostar phases has been observed toward MC27/L1521F \citep{favre2020}. This source hosts a very-low-luminosity object (VeLLO) --- objects with a luminosity of $L_{int} < 0.1 L_{\astrosun}$ associated with dense cores \citep{https://doi.org/10.1002/asna.200510446} --- where the embedded protostar is just beginning to heat up \citep{bourke2006}. \citet{burkhardt2021a} detected the single-ring benzonitrile in MC27/L1521F, indicating its survival during the onset of star formation.

\begin{figure*}[htb!]
    \centering
    \begin{minipage}{\textwidth}
    \centering
    \raisebox{-0.5\height}{\includegraphics[height=1in]{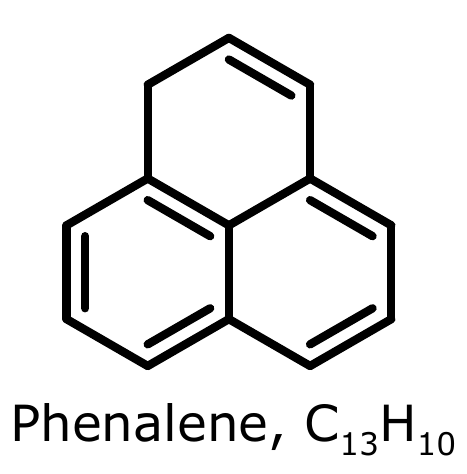}}
    \hfill
    \raisebox{-0.5\height}{\includegraphics[height=1.725in]{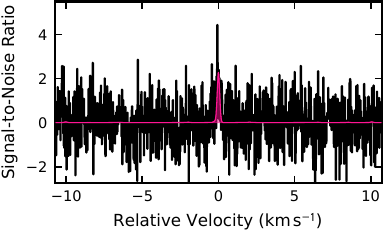}}
    \hfill
    \raisebox{-0.5\height}{\includegraphics[height=1.725in]{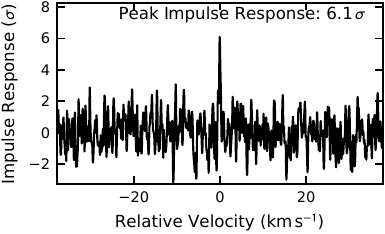}}
    \end{minipage}
    \caption{(Left) Structure of 1H-phenalene (\ce{C13H10}). (Center) Velocity-stacked spectra and (right) matched filter response of phenalene emission in the ARKHAM data toward MC27/L1521F generated using the methodologies outlined in \citet{loomis2021}.
    }
    \label{fig:PhenMC27}
\end{figure*}

The three-ring PAH 1H-phenalene (\ce{C13H10}), hereafter phenalene, has been previously detected in TMC-1 CP \citep{cabezas2025}. Phenalene consists of a naphthalene subunit, with an ortho- and peri-fused cyclohexene ring (see Figure \ref{fig:PhenMC27}). An H atom from the \ce{CH2} group in phenalene can be readily removed to form a resonantly stabilized radical, phenalenyl. Its relevance to combustion chemistry motivated the first gas-phase laboratory detection of phenalenyl \citep{oconnor2011}, and it has since been speculated to play an important role in PAH growth in flames \citep{johansson2018}.

Here, we report the detection of phenalene toward MC27/L1521F in high abundance, the first three-ring PAH detected outside of TMC-1 CP. 

\section{Observations}

All observations were acquired on the 100\,m Robert C. Byrd Green Bank Telescope (GBT) as part of two large, ongoing spectral line surveys: GBT Observations of TMC-1: Hunting for Aromatic Molecules (GOTHAM;~\citealt{mcguire2020}) and A Rigorous $K$/$Ka$-Band Hunt for Aromatic Molecules (ARKHAM;~\citealt{burkhardt2021a}). Details of the observing strategies and data reduction procedures were previously described by \citet{xue2025}, including advances in artifact and radio-frequency interference removal, atmospheric opacity corrections, and baseline removal. 

Briefly, the GOTHAM survey targets the ``Cyanopolyyne Peak'' of TMC-1 centered at $\alpha_\mathrm{J2000}$ = 04\fh41\fm42.5\fs, $\delta_\mathrm{J2000}$ = +25\arcdeg41\arcmin26.8\arcsec~and spans 29\,GHz of bandwidth between 4--36\,GHz at a spectral resolution of 1.4\,kHz. The root-mean-squared (RMS) noise level ranges between 3.8--15.0\,mK across the majority of the observed frequency range. Flux calibration was carried out using switched noise-diode measurements, yielding an estimated antenna temperature accuracy of 10--20\,\%. The GOTHAM observations cover the $C$-, $X$-, $Ku$-, $K$- and $Ka$-bands. The ARKHAM survey was designed to target multiple sources for benzonitrile detection. In particular, MC27/L1521F was observed centering the GBT at $\alpha_\mathrm{J2000}$ = 04\fh28\fm39.3\fs, $\delta_\mathrm{J2000}$ = +26\arcdeg51\arcmin39.0\arcsec covering a frequency range of 22--30.5\,GHz of the $K$- and $Ka$-bands. The ARKHAM data were recalibrated using the newly developed \texttt{GOTHAM Spectral Pipeline} \citep{xue2025} to properly correct for zenith opacity, particularly around the 22\,GHz water vapor line, and to better constrain the noise properties by reducing the noise covariance. The RMS noise level ranges between 5.2--23.5\,mK across the majority of the observed frequency range.

\section{Astronomical Analysis}

The laboratory rotational spectrum of phenalene was measured and fit by \citet{cabezas2025} and its spectroscopic constants, partition function, and spectral line catalog are publicly available in the Cologne Database for Molecular Spectroscopy (CDMS)\footnote{\url{https://cdms.astro.uni-koeln.de/classic/}} \citep{endres2016}. Together with its laboratory analysis, \citet{cabezas2025} also reported the first interstellar detection of phenalene in the QUIJOTE observations conducted with the 40\,m Yebes radio telescope toward TMC-1 CP. Simulating the radio emission of phenalene under TMC-1 CP conditions (see Tables~\ref{tab:MCMCpriors} and \ref{tab:mcmcTMC-1}) and performing a velocity-stack and matched filter response analysis \citep{loomis2021}, we also detect this molecule in the fifth data reduction (DRV) of the GOTHAM dataset \citep{xue2025} from the GBT with a significance of 18.2\,$\sigma$ (see Figure~\ref{fig:PhenTMC-1}). Using the Markov Chain Monte Carlo (MCMC) analysis adopted from previous work~\citep{mcguire2021,loomis2021}, we derive an excitation temperature of $9.91^{+1.73}_{-1.33}$\,K and a column density of $1.53^{+0.27}_{-0.14}\times 10^{13}\,\mathrm{cm^{-2}}$ that best describe the emission from phenalene in TMC-1 CP. This is in agreement with the previously reported temperature of $(7.9 \pm 1.2)\,$K and column density of $(2.8 \pm 1.6) \times 10^{13}\,\mathrm{cm^{-2}}$ by \citet{cabezas2025}.

The search for phenalene toward MC27/L1521F was performed analogously. We first simulated the rotational spectrum of benzonitrile under the physical conditions present in MC27/L1521F and performed an MCMC analysis on the newly reduced ARKHAM data informed by the previous detection of benzonitrile in this source \citep{burkhardt2021a}. The results are listed in Table~\ref{tab:mcmc} and agree with the previous analysis. We performed an MCMC analysis using priors constrained by our benzonitrile detection in MC27/L1521F to derive physical parameters that best describe the emission of phenalene (see Tables~\ref{tab:MCMCpriors} and \ref{tab:mcmc}).
Using these values, we simulated the phenalene emission in MC27/L1521F and carried out a velocity-stack and matched filter analysis (see Figure~\ref{fig:PhenMC27}) of the ARKHAM data \citep{burkhardt2021a}. We derive a detection significance of phenalene toward MC27/L1521F of 6.1$\,\sigma$. At the RMS noise level of the observations of 5.2--23.5\,mK, individual emission lines of phenalene are not detected above the noise. From the MCMC analysis, we derive an excitation temperature of $4.91^{+0.08}_{-0.08}\,$K and a column density of $1.47^{+0.49}_{-0.35} \times 10^{13}\,\mathrm{cm^{-2}}$, which is similar to the column density of phenalene found in TMC-1 CP. 

\begin{figure}[htb!]
    \centering
    \includegraphics[width=\linewidth]{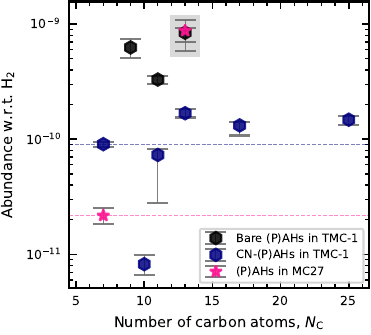}
    \caption{Abundance of detected (polycyclic) aromatic hydrocarbons ((P)AHs) in TMC-1 CP (blue and black hexagons) and in MC27/L1521F (magenta stars) with respect to the molecular hydrogen abundance in each source and as a function of number of carbon atoms per molecule, $N_\mathrm{C}$. Dashed horizontal lines are the propagated benzonitrile abundances to guide the eye. The area shaded in grey highlights the high abundance of phenalene in TMC-1 and MC27/L1521F. Column density values and the references they originate from are reported in Table~\ref{tab:CDcomp}.}
    \label{fig:abundance}
\end{figure}

The ratio of phenalene-to-benzonitrile, however, differs significantly between both sources. The phenalene/benzonitrile ratio in TMC-1 CP is ${\sim}9$ (9.27$^{+1.71}_{-0.91}$), whereas this ratio is ${\sim}40$ (40.2$^{+14.7}_{-11.5}$) in MC27/L1521F, making phenalene a factor of ${\sim}$4 more abundant in MC27/L1521F when comparing to benzonitrile than in TMC-1 CP. Figure~\ref{fig:abundance} compiles the abundances of interstellar (P)AHs with respect to molecular hydrogen (values listed in Table~\ref{tab:CDcomp}) as a function of molecular complexity, starting with the single-ring benzonitrile. While generally the bare (unsubstituted) PAHs in TMC-1 CP are approximately one order of magnitude more abundant than CN-(P)AHs, this ratio is clearly changed in MC27/L1521F. 

\section{Discussion}

To date, the formation pathways of most PAHs found at the earliest stages of star formation, in sources such as TMC-1 or MC27/L1521F, are predominantly unknown. One of the major known formation routes of PAHs is the \ce{H}-abstraction acetylene (\ce{C2H2}) addition (HACA) mechanism that involves substantial reaction barriers \citep{frenklach1985,reizer2022}. Such barriers cannot be surmounted in cold cores, but can be overcome in high-temperature environments such as those found in flames or combustion processes or in carbon-rich circumstellar environments \citep{frenklach2024}. In such environments, the production of the phenalenyl radical has been studied through a reaction scheme of acenaphthylene and methane in an electrical discharge \citep{levey2022}. This was supported by prior theoretical work that characterized a barrierless formation route to the phenalenyl radical via phenalene in the reaction of 1-acenaphthyl and methyl radicals \citep{porfiriev2020}. If occurring in tandem with, e.g., HACA, this process might produce larger PAHs in flames and the ISM. \citet{zhao2020} proposed another gas-phase formation route to phenalene via the reaction of the 1-naphthyl radical with methylacetylene and allene under high-temperature conditions, which likewise could occur in circumstellar environments. 

MC27/L1521F is a source hosting a newly formed Class 0 protostar \citep{tokuda2026} producing ultraviolet (UV) radiation. VeLLOs are thought to undergo episodic accretion episodes, in which the accretion rate may be substantially enhanced relative to their present quiescent state. As the far-UV (FUV) field scales with accretion rate \citep{yang2011}, MC27/L1521F may have experienced periods with a significantly enhanced FUV field relative to both the standard interstellar radiation field, as well as TMC-1 CP. For accretion rates on the order of 10$^{-8}$ $M_{\odot}$ yr$^{-1}$, local radiation fields of G$_0$ $\sim$10$^2$ -- 10$^3$ are plausible at radii of $\sim$100 AU \citep{bergin2003}.

The enhanced phenalene/benzonitrile ratio in MC27/L1521F may reflect differing photostabilities of the two species. With a more intense UV radiation field present in MC27/L1521F, photodissociation of benzonitrile could be more efficient (see, for example, \citet{kamer2023,rap2024,debes2025}) compared to that of phenalene, and thus enhance the phenalene/benzonitrile ratio. However, absolute interstellar photodissociation rates for the neutral species have not been reported, making quantitative comparisons difficult. The exceptionally weak C-H bonds in the \ce{CH2} moiety of phenalene (258.1\,kJ\,mol$^{-1}$; \citealt{oconnor2017}) likely lowers its dissociation thresholds compared to fully aromatic PAHs, which may enhance fragmentation \citep{porfiriev2020,marciniak2021}. Conversely, phenalene possesses a higher density of states than benzonitrile, which may enhance internal energy redistribution and cooling, resulting in a longer lifetime prior to fragmentation. In addition, phenalene has a larger number of vibrational modes from which infrared emission can occur. The combination of these factors would tend to favor stabilization of phenalene as compared to benzonitrile.

Recent work has shown that recurrent fluorescence (RF) offers an efficient route for radiatively cooling PAH cations, quenching dissociation and leading to their enhanced stability upon UV excitation \citep{stockett_efficient_2023,subramani2025,bull2025}. Although RF rates for the phenalene cation have not yet been measured, its larger vibrational manifold may favor more efficient radiative stabilization compared to the benzonitrile cation. The relative stability of phenalene and benzonitrile under interstellar UV irradiation depends sensitively on the competition between fragmentation and radiative stabilization. 

The abundance (with respect to H$_2$) that we derive for benzonitrile in MC27/L1521F is roughly five times lower than that in TMC-1 CP, ${\sim}2\times10^{-11}$ versus ${\sim}9\times10^{-11}$, respectively. In contrast, the phenalene abundance in the two sources is similar, ${\sim}10^{-9}$. However, it is important to note that there is considerable uncertainty on the H$_2$ column density reported for MC27/L1521F. The H$_2$ column density may be higher within the primary beam of the GBT than the weighted mean value originally derived by \citet{chitsazzadeh2014}, resulting in lower abundances of both benzonitrile and phenalene. 

Benzonitrile has now been observed in a number of cold molecular cloud sources in Lupus, Taurus, Aquila, and Serpens \citep{mcguire2018a, burkhardt2021a,agundez2023}, as well as recently in warmer gas in the Central Molecular Zone \citep{rivilla2026}. The sources containing protostars, L1527 in Taurus and L483 in Aquila, have the lowest benzonitrile abundances. Variations in the abundances across the sources may reflect differences in the physical conditions, such as the local FUV fields, or chemical evolution effects, such as differences in the partitioning of carbon and oxygen between the gas and dust. MC27/L1521F is a more dynamically evolved and denser core than TMC-1 CP, exhibiting strong CO freezeout \citep{crapsi2004}; therefore, H$_3^+$ and N$_2$H$^+$ (which are predominantly destroyed through reactions with gas-phase CO) will likely be more abundant. Such differences in key molecular ions may cause downstream chemical effects on the carbon reservoirs, resulting in differences in the phenalene/benzonitrile abundance ratios.

Lastly, phenalene has been noted to potentially form pyrene \citep{frenklach2024}, another interstellar PAH that has been found in TMC-1 CP via its CN-derivatives \citep{wenzel2024,wenzel2025}. So far, the synthesis route of pyrene in the cold ISM is not understood; however, \citet{zhao2018} showed its formation via a HACA mechanism from phenanthrenyl in the high-temperature environments of carbon-rich circumstellar envelopes. Detecting the three-ring PAH, phenalene, in the cold cloud MC27/L1521F shows the critical importance of understanding interstellar PAH chemistry, as these large carbon reservoirs will be directly available to star and planet formation.

The unsubstituted phenalene is another candidate to test the so-called ``proxy method'', in which column densities of some ``bare'' PAHs have been inferred through the detection of cyano-substituted PAHs with large permanent electric dipole moments. To date, only one bare aromatic molecule and one of its cyano-functionalized derivatives have both been detected in the same source, namely indene and 2-cyanoindene in TMC-1 CP \citep{burkhardt2021,cernicharo2021,sita2022,xue2025}, with a ratio of parent-to-child of H/CN = 76$^{+49}_{-36}$ based on the column densities in \citet{xue2025}. The detection of the untagged parent phenalene presents another potential opportunity to better constrain the proxy method. As such, we carried out preliminary work (see Appendix~\ref{sec:calcs}) that verifies CN addition followed by H atom elimination at sites 2--9 of phenalene occurs over deeply submerged barriers. Combining this with overall rate coefficients estimated using classical capture theory rate coefficients and the modeled rate dependence of CN tagged child-to-bare-parent-PAH determined in \citet{wenzel2025} and used in \citet{wenzel2025c} yields 2.18$^{+2.96}_{-1.19}$$\times$ 10$^{12}$, $\mathrm{cm^{-2}}$, with the expected isomer specific abundance being  2.72$^{+3.70}_{-1.49}$ $\times$ 10$^{11}$ $\mathrm{cm^{-2}}$, suggesting that 1-cyano-1H-phenalene could be formed with an abundance up to 9.07$^{+12.34}_{-4.98}$ $\times$ 10$^{11}$ $\mathrm{cm^{-2}}$. However, a dedicated interstellar search for the cyanophenalenes will require their rotational spectra to be first characterized in the laboratory. 

\section{Conclusions}

We report the detection of a three-ring PAH outside of TMC-1 CP. Phenalene (\ce{C13H10}) was identified using legacy ARKHAM data \citep{burkhardt2021a} toward the dense molecular cloud core MC27/L1521F with a column density of $1.47^{+0.49}_{-0.35}\times 10^{13}\,\mathrm{cm^{-2}}$ at an excitation temperature of $4.91^{+0.08}_{-0.08}$\,K. The phenalene/benzonitrile ratio is four times larger in MC27/L1521F than the one derived for TMC-1 CP. So far, no low-temperature formation pathways to phenalene that could occur under cold cloud conditions are known. This motivates further studies of UV-driven PAH chemistry under interstellar conditions. Nevertheless, this detection signifies the prevalence of PAHs across the cold ISM, where they are carbon reservoirs available during protostar formation.

\section{Data access \& code}
    The raw data of the GOTHAM observations are publicly available from \citet{xue2025} and in the GBT Legacy Data Archive\footnote{\footnotesize \url{https://greenbankobservatory.org/portal/gbt/gbt-legacy-archive/gotham-data/}}. The ARKHAM data is available in \citet{burkhardt2021a}. The \texttt{GOTHAM Spectral Pipeline}\footnote{\url{https://github.com/cixue/gotham-spectral-pipeline}} used for the astronomical data reduction is publicly available in \citet{gotham-spectral-pipeline} and the analysis was carried out using the \texttt{molsim}\footnote{\url{https://github.com/bmcguir2/molsim}} open-source software package; an archival version of the code can be accessed at \citet{molsim2024}.

\begin{acknowledgements}
We gratefully acknowledge support from NSF grants AST-1908576, AST-2205126, and AST-2307137. G.W. and B.A.M. acknowledge the support of the Arnold and Mabel Beckman Foundation Beckman Young Investigator Award. G.W. and B.A.M. also gratefully acknowledge the support of Schmidt Family Futures. I.R.C. acknowledges support from the University of British Columbia and the Natural Sciences and Engineering Research Council of Canada (RGPIN-2022-04684). I.R.C. and T.H.S. acknowledge the support of the Canadian Space Agency through grant 24AO3UBC14 and this work was in part supported through the computational resources and services provided by Advanced Research Computing at the University of British Columbia. C.X. acknowledges support from the National Science Foundation under Cooperative Agreement 2421782 and the Simons Foundation grant MPS-AI-00010515 awarded to the NSF-Simons AI Institute for Cosmic Origins -- CosmicAI, https://www.cosmicai.org/. The National Radio Astronomy Observatory is a facility of the National Science Foundation operated under cooperative agreement by Associated Universities, Inc. The Green Bank Observatory is a facility of the National Science Foundation operated under cooperative agreement by Associated Universities, Inc.  
\end{acknowledgements}

\begin{contribution}

G.W.: Conceptualization, Data Curation, Formal analysis, Investigation, Visualization, Writing – original draft. T.H.S.: Data Curation, Investigation, Resources, Writing – original draft. C.X.: Data Curation, Investigation, Software, Validation. E.A.B., A.M.B, M.A.C, M.D., Z.T.P.F, A.L., C.N.S., R.H.J.W.: Investigation. A.J.R.: Resources, Supervision. M.C.M.: Funding acquisition, Resources, Supervision. B.A.M.: Funding acquisition, Methodology, Resources, Supervision. I.R.C.: Funding acquisition, Investigation, Resources, Supervision, Writing – original draft. All authors contributed to Writing – review \& editing and are members of the GOTHAM collaboration.

\end{contribution}

\facility{GBT}

\software{
    \texttt{GOTHAM Spectral Pipeline} \citep{xue2025}, 
    \texttt{molsim} \citep{molsim2024}
    }

\appendix
\restartappendixnumbering

\section{MCMC Analyses of Phenalene in TMC-1 and MC27/L1521F}

\begin{figure*}[htb!]
    \centering
    \begin{minipage}{\textwidth}
    \centering
    \raisebox{-0.5\height}{\includegraphics[height=1.25in]{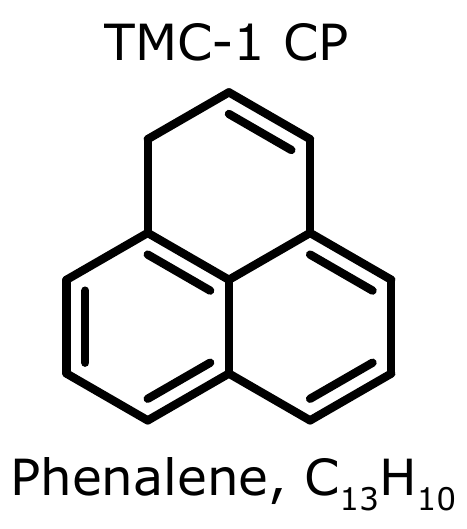}}
    \hfill
    \raisebox{-0.5\height}{\includegraphics[height=1.725in]{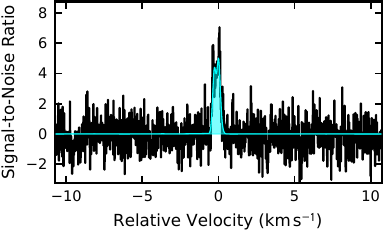}}
    \hfill
    \raisebox{-0.5\height}{\includegraphics[height=1.725in]{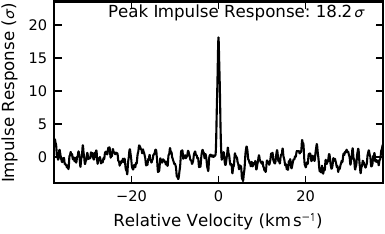}}
    \end{minipage}\\
    \begin{minipage}{\textwidth}
    \centering
    \raisebox{-0.5\height}{\includegraphics[height=1.25in]{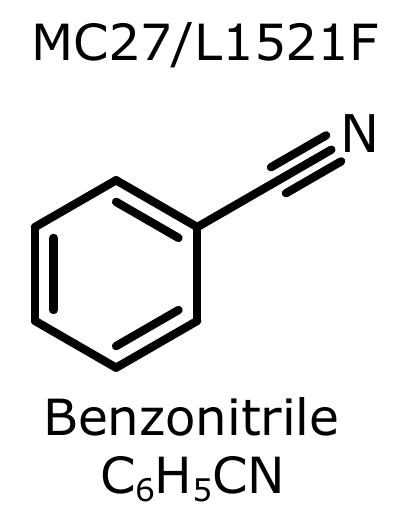}}
    \hspace{0.25cm}
    \hfill
    \raisebox{-0.5\height}{\includegraphics[height=1.725in]{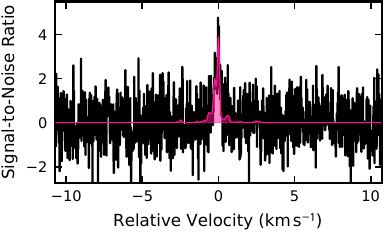}}
    \hfill
    \raisebox{-0.5\height}{\includegraphics[height=1.725in]{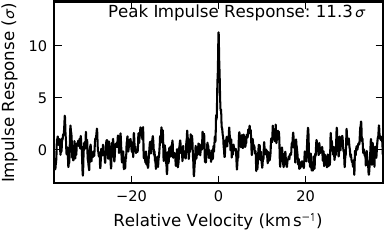}}
    \end{minipage} \\
    \caption{(Top left) Velocity-stacked spectra of the simulated phenalene emission (aqua) and observational data (black) and (top right) matched filter response of phenalene in the GOTHAM data toward TMC-1 CP. (Bottom left) Velocity-stacked spectra of the simulated benzonitrile emission (magenta) and observational data (black) and (bottom right) matched filter response of benzonitrile in the ARKHAM data toward MC27/L1521F. All plots were generated using the methodologies outlined in \citet{loomis2021}.
    }
    \label{fig:PhenTMC-1}
\end{figure*}

\begin{figure*}[htb!]
\includegraphics[width=\textwidth]{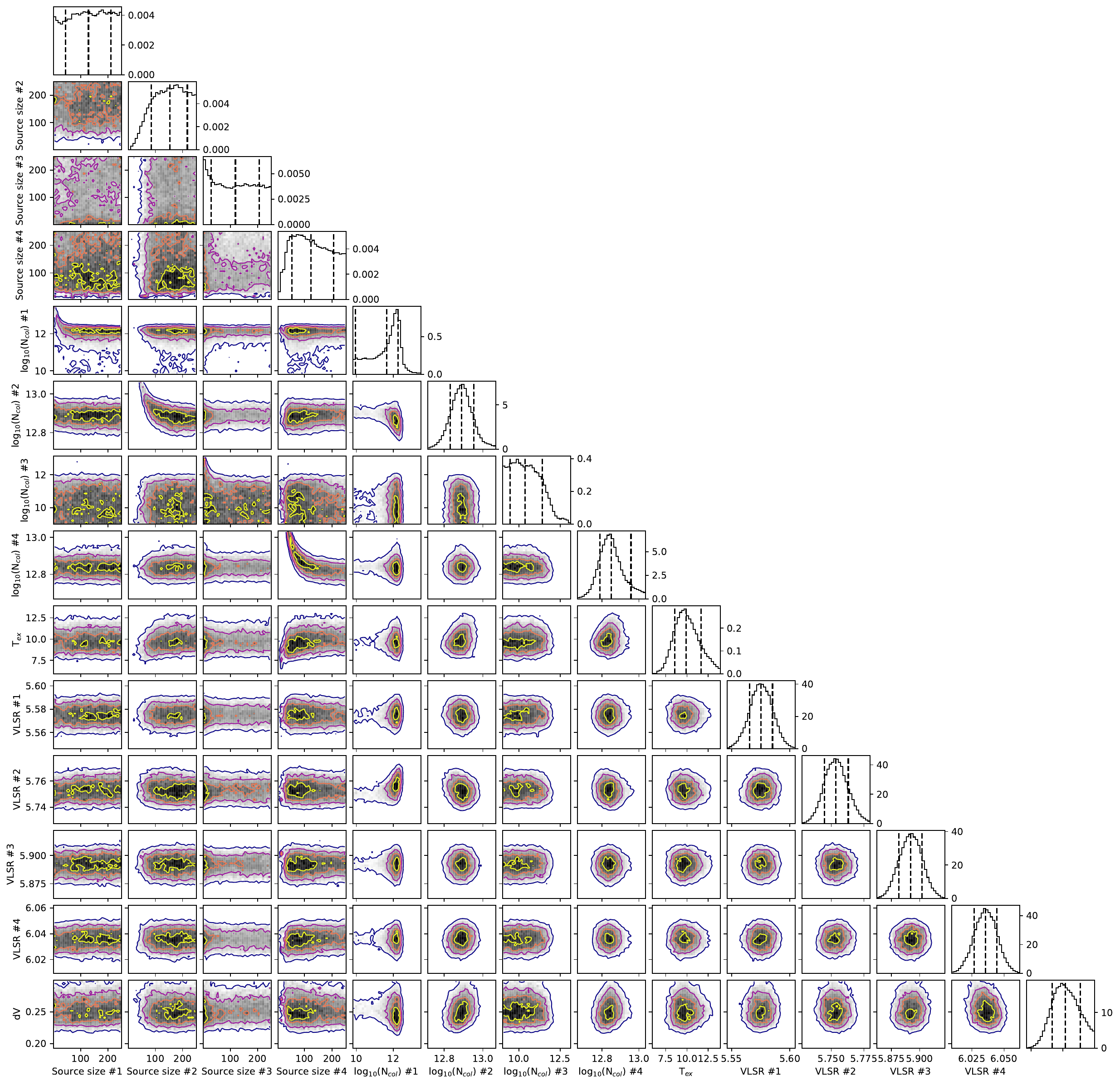}
    \caption{Parameter covariances and marginalized posterior distributions for the MCMC fit of phenalene in TMC-1 CP. 16$^\mathrm{th}$, 50$^\mathrm{th}$, and 84$^\mathrm{th}$ confidence intervals (corresponding to $\pm1\,\sigma$ for a Gaussian posterior distribution) are depicted as vertical lines on the diagonal plots.}
    \label{fig:cornerplot_tmc-1}
\end{figure*}

\begin{figure*}[htb!]
\centering
\includegraphics[width=0.495\textwidth]{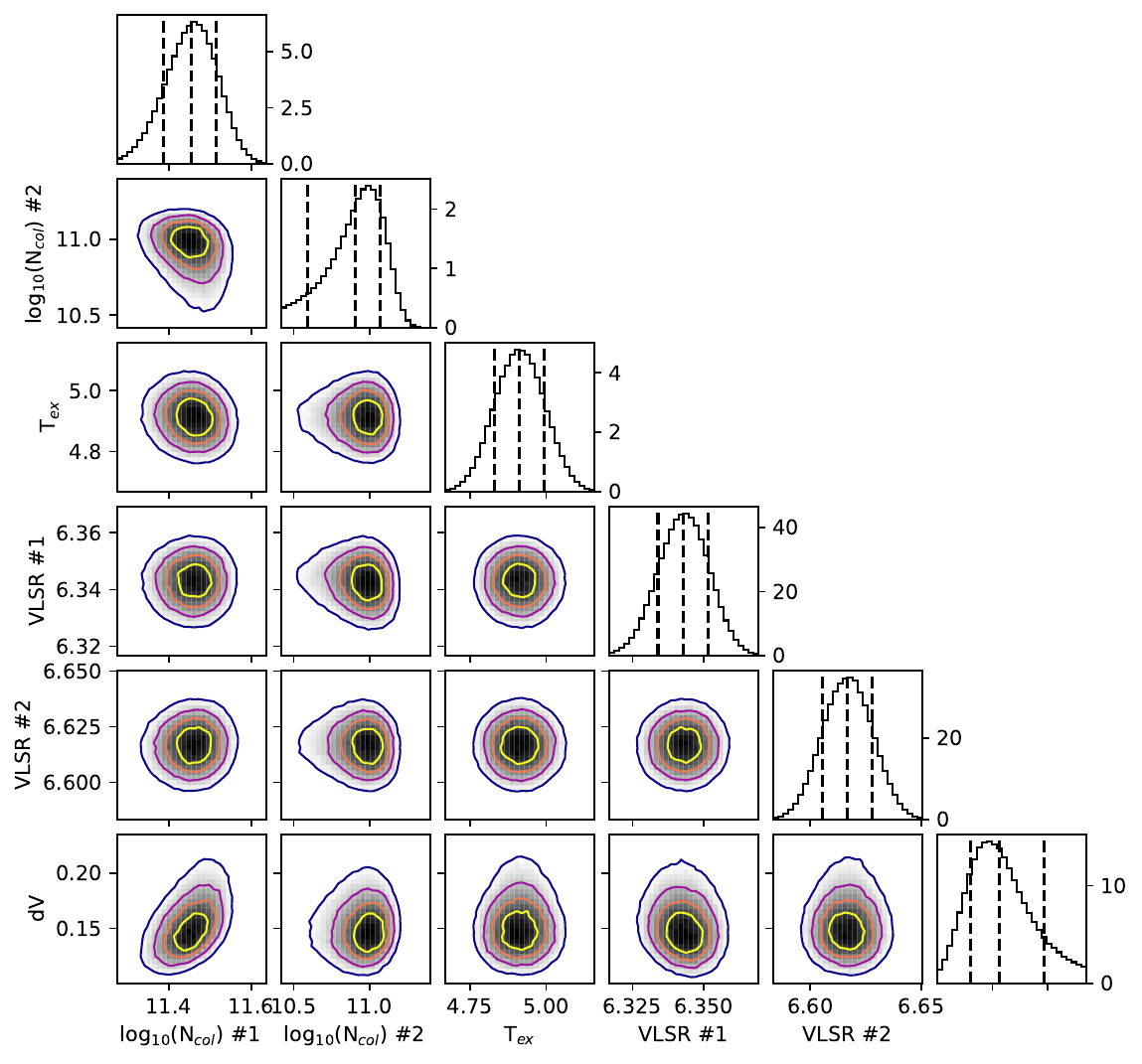} \hfill
\includegraphics[width=0.495\textwidth]{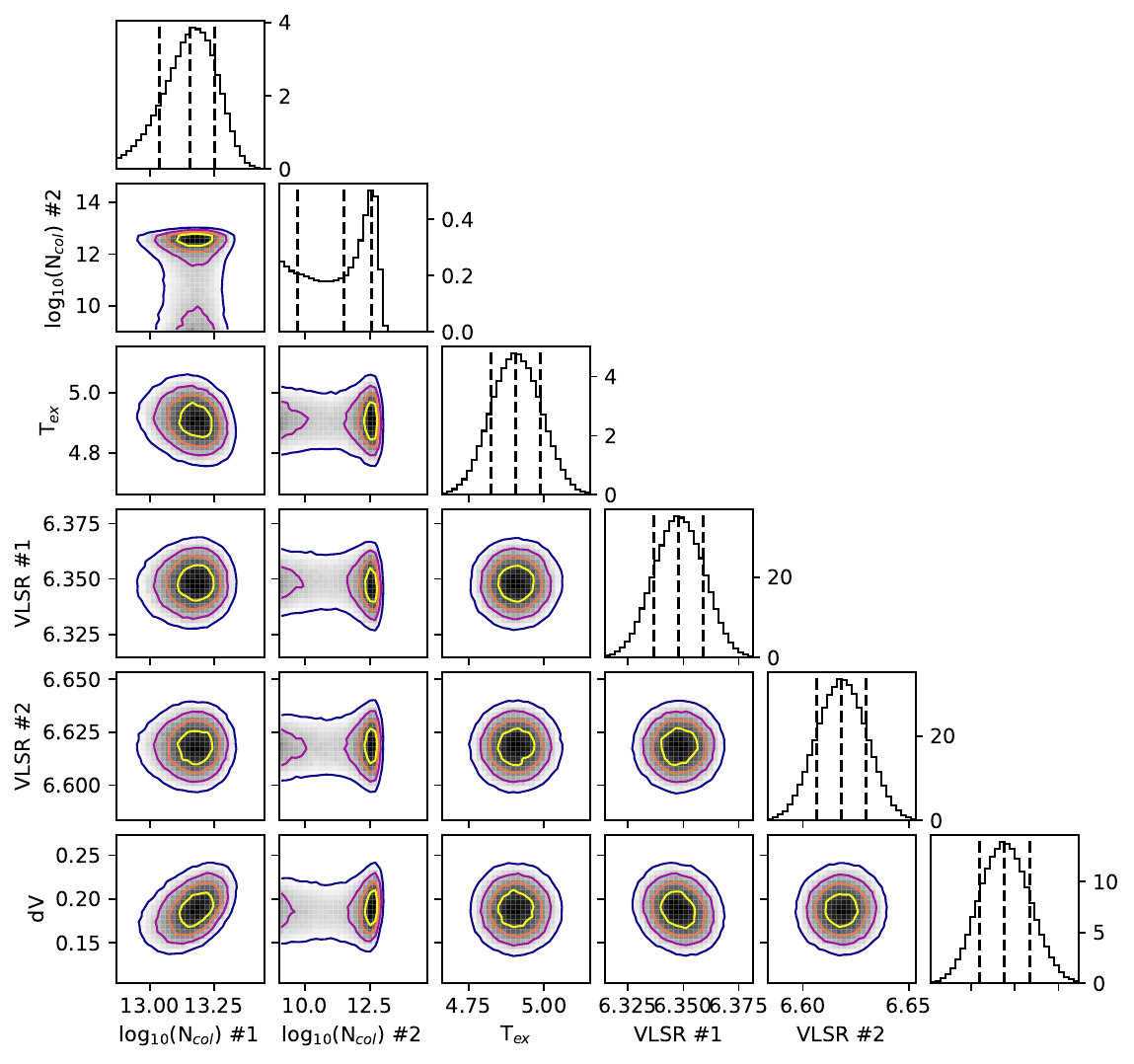}
    \caption{Parameter covariances and marginalized posterior distributions for the MCMC fits of benzonitrile (left) and phenalene (right) in MC27/L1521F. 16$^\mathrm{th}$, 50$^\mathrm{th}$, and 84$^\mathrm{th}$ confidence intervals (corresponding to $\pm1\,\sigma$ for a Gaussian posterior distribution) are depicted as vertical lines on the diagonal plots.}
    \label{fig:cornerplot_mc27_benzonitrile}
\end{figure*}

MCMC analyses were performed analogously to the methodologies outlined by \citet{loomis2021} and priors were informed by previous works of the GOTHAM team and benzonitrile detections from the ARKHAM project, see in particular \citet{xue2025} and \citet{burkhardt2021a}. The priors used in this work for MCMC analyses of molecular emission in TMC-1 CP and MC27/L1521F are listed in Table~\ref{tab:MCMCpriors} and the results are reported in Tables~\ref{tab:mcmcTMC-1} and \ref{tab:mcmc} for molecular emission in TMC-1 CP and MC27/L1521F, respectively. Figure~\ref{fig:PhenTMC-1} demonstrates the velocity-stack and matched filter analysis of phenalene in TMC-1 CP (upper panels) and benzonitrile in MC27/L1521F (lower panels). Note that the weak signal observed at non-zero relative velocities in the benzonitrile velocity-stack is from  benzonitrile. Due to $^{14}$N nuclear quadrupole hyperfine splitting, benzonitrile contains many closely spaced lines that will be contained in the extracted spectral windows centered on each line. The posteriors of the benzonitrile detection in MC27/L1521F were used to inform the priors used for the phenalene analysis in the same source. Figures~\ref{fig:cornerplot_tmc-1} and \ref{fig:cornerplot_mc27_benzonitrile} depict the covariance plots of each detection.

\begin{table*}[htb]
    \caption{Priors used for the MCMC analysis. A Gaussian distribution, $\mathcal{N}(\mu,\sigma^2)$, with mean $\mu$ and variance $\sigma^2$, was used for velocity, $v_\mathrm{LSR}$, both in TMC-1 and in MC27/L1521F. For TMC-1, a uniform (unweighted) distribution, $\mathcal{U}\{\mathrm{min},\mathrm{max}\}$, between $\mathrm{min}$ and $\mathrm{max}$ was used for the source size, column density, $N_\mathrm{T}$, excitation temperature, $T_\mathrm{ex}$, and linewidth, $\Delta V$. Due to phenalene's faint emission in MC27/L1521F, priors were more constrained with a fixed source size, $\delta(x)$, and Gaussian distributions for excitation temperature, $T_\mathrm{ex}$, and linewidth, $\Delta V$. The column density was varied using a uniform (unweighted) distribution, $\mathcal{U}\{\mathrm{min},\mathrm{max}\}$. Priors were adopted from our previous comprehensive analysis of aromatic molecules in TMC-1 \textbf{using four velocity components} \citep{xue2025} and benzonitrile in MC27/L1521F \textbf{using two velocity components} \citep{burkhardt2021a}.}
    \centering
    \begin{tabular}{cccccc}
    \\
    \toprule
    \multicolumn{6}{c}{\textbf{Priors for Aromatic Molecules in TMC-1 CP}}\\
    \midrule
      Component &  $v_\mathrm{LSR}$	&	Size	&	$\mathrm{log_{10}}(N_\mathrm{T})$	&	$T_\mathrm{ex}$	&	$\Delta V$	\\
	No. & ($\mathrm{km\,s^{-1}}$) &($^{\prime\prime}$)	&	($\mathrm{cm}^{-2}$)	&	($\mathrm{K}$)	&	($\mathrm{km\,s^{-1}}$)\\
	\midrule
	1 & {$\mathcal{N}(5.575,0.01)$} & \multirow{4}{*}{$\mathcal{U}\{\mathrm{\mathrm{min,max}}\}$}	 &  	 \multirow{4}{*}
 {$\mathcal{U}\{\mathrm{min,max}\}$} &
 \multirow{4}{*}{$\mathcal{U}\{\mathrm{\mathrm{min,max}}\}$}	 & 	 \multirow{4}{*}{$\mathcal{U}\{\mathrm{\mathrm{min,max}}\}$}\\
	2 &	{$\mathcal{N}(5.767,0.01)$} &	 & 	 & 	 & 	 \\
        3 & {$\mathcal{N}(5.892,0.01)$} &	 & 	 &   & 	 \\
	4 &	{$\mathcal{N}(6.018,0.01)$} &  &  & 	 & 	 \\
 \midrule
        Min & $0.0$ & $1$ & $9.0$ & $3.0$ & $0.1$ \\
        Max & $10.0$ & $250$ & $14.0$ & $15.0$ & $0.3$ \\
\midrule
 \multicolumn{6}{c}{\textbf{Priors for Aromatic Molecules in MC27/L1521F}} \\
            \midrule
      Component &  $v_\mathrm{LSR}$	&	Size	&	$\mathrm{log_{10}}(N_\mathrm{T})$	&	$T_\mathrm{ex}$	&	$\Delta V$	\\
	No. & ($\mathrm{km\,s^{-1}}$) &($^{\prime\prime}$)	&	($\mathrm{cm}^{-2}$)	&	($\mathrm{K}$)	&	($\mathrm{km\,s^{-1}}$)\\
	\midrule
	1 & {$\mathcal{N}(6.341,0.01)$} & \multirow{2}{*}{$\delta(400)$}	 &  	 \multirow{2}{*}
 {$\mathcal{U}\{{\mathrm{min,max}}\}$} &
 \multirow{2}{*}{$\mathcal{N}(4.91,0.10)$}	 & 	 \multirow{2}{*}{$\mathcal{N}(0.16,0.04)$}\\
	2 &	{$\mathcal{N}(6.618,0.01)$} &	 & 	 & 	 & 	 \\
 \midrule
        Min & $0.0$ &  & $10.0$ & $1.0$ & $0.1$ \\
        Max & $10.0$ &  & $15.0$ & $9.0$ & $0.4$ \\
	\bottomrule
\\
    \end{tabular}\\
    \label{tab:MCMCpriors}
\end{table*}

\begin{table*}[hbt!]
    \centering
    \caption{Summary statistics of the marginalized posterior probability distributions from the MCMC analysis for phenalene in TMC-1. Priors used are reported in Table~\ref{tab:MCMCpriors}. The uncertainties are represented by the 16$^\mathrm{th}$ and 84$^\mathrm{th}$ percentile, also known as the $68\,\%$ confidence interval, which corresponds to 1$\,\sigma$ for a Gaussian distribution. The total column density and its uncertainty are derived by marginalizing over the posterior distributions for the column densities of each individual component and reporting the 50$^\mathrm{th}$, 16$^\mathrm{th}$, and 84$^\mathrm{th}$ percentiles.}
    \begin{tabular}{cccccc}
    \\
    \toprule
    \multicolumn{6}{c}{\textbf{Phenalene (\ce{C13H10}) in TMC-1 CP}}\\
\midrule

Component & $v_\mathrm{LSR}$	&	Size	&	$N_\mathrm{T}$	&	$T_\mathrm{ex}$	&	$\Delta V$	\\
No. & {($\mathrm{km\,s^{-1}}$)} & ($^{\prime\prime}$)	&	($10^{12}\,\mathrm{cm}^{-2}$)	&	($\mathrm{K}$)	&	($\mathrm{km\,s^{-1}}$)\\
	\midrule

        1 & $5.575^{+0.010}_{-0.010}$	 & 	$128^{+82}_{-84}$	 & 	$0.44^{+1.37}_{-0.43}$	 & 	 \multirow{4}{*}{$9.91^{+1.73}_{-1.33}$}	 & 	 \multirow{4}{*}{$0.254^{+0.024}_{-0.020}$}\\
		2 & $5.754^{+0.009}_{-0.009}$	 & 	$153^{+63}_{-68}$	 & 	$7.77^{+1.20}_{-0.98}$	 & 	 & 	 \\		
        3 & $5.892^{+0.010}_{-0.010}$	 & 	$118^{+88}_{-89}$	 & 	$0.02^{+0.23}_{-0.02}$	 & 	 & 	 \\
		4 & $6.036^{+0.009}_{-0.009}$	 & 	$123^{+82}_{-68}$	 & 	$7.08^{+1.97}_{-0.93}$	 & 	 & 	 \\
	\midrule
		\multicolumn{6}{c}{$N_\mathrm{T}$(Total): $1.53^{+0.27}_{-0.14}\times 10^{13}$\,cm$^{-2}$}\\
  \bottomrule
    \end{tabular}
    \label{tab:mcmcTMC-1}
\end{table*}

\begin{table*}[hbt!]
    \centering
    \caption{Same as Table~\ref{tab:mcmcTMC-1} for 
    benzonitrile and phenalene in MC27/L1521F.}
    \begin{tabular}{cccccc}
    \\
    \toprule

    \multicolumn{6}{c}{\textbf{Benzonitrile (\ce{C6H5CN}) in MC27/L1521F}}\\
\midrule
      Component & $v_\mathrm{LSR}$	&	Size	&	$N_\mathrm{T}$	&	$T_\mathrm{ex}$	&	$\Delta V$	\\
	No. & {($\mathrm{km\,s^{-1}}$)} & ($^{\prime\prime}$)	&	($10^{11}\,\mathrm{cm}^{-2}$)	&	($\mathrm{K}$)	&	($\mathrm{km\,s^{-1}}$)\\
	\midrule 
1 & $6.343^{+0.009}_{-0.009}$ &
$[400]$ &
$2.85^{+0.42}_{-0.41}$ & \multirow{2}{*}{$4.91^{+0.08}_{-0.08}$} & 	 \multirow{2}{*}{$0.156^{+0.041}_{-0.026}$} \\ 
2 & $6.617^{+0.011}_{-0.011}$ &
$[400]$ &
$0.81^{+0.37}_{-0.42}$ & & \\

	\midrule

		\multicolumn{6}{c}{$N_\mathrm{T}$(Total): $3.66^{+0.56}_{-0.58} \times 10^{11}$\,cm$^{-2}$}\\

\toprule
    \multicolumn{6}{c}{\textbf{Phenalene (\ce{C13H10}) in MC27/L1521F}}\\
    \midrule
      Component & $v_\mathrm{LSR}$	&	Size	&	$N_\mathrm{T}$	&	$T_\mathrm{ex}$	&	$\Delta V$	\\
	No. & {($\mathrm{km\,s^{-1}}$)} & ($^{\prime\prime}$)	&	($10^{13}\,\mathrm{cm}^{-2}$)	&	($\mathrm{K}$)	&	($\mathrm{km\,s^{-1}}$)\\
    \midrule
1 & $6.348^{+0.011}_{-0.011}$ &
$[400]$ &
$1.44^{+0.36}_{-0.35}$ & \multirow{2}{*}{$4.91^{+0.08}_{-0.08}$} & \multirow{2}{*}{$0.188^{+0.029}_{-0.028}$} \\
2 & $6.618^{+0.012}_{-0.012}$ &
$[400]$ &
$0.03^{+0.34}_{-0.03}$ & & \\

	\midrule

		\multicolumn{6}{c}{$N_\mathrm{T}$(Total): $1.47^{+0.49}_{-0.35} \times 10^{13}$\,cm$^{-2}$}\\

  \bottomrule
    \end{tabular}
    \label{tab:mcmc}
\end{table*}

\section{Column Densities of Cyclic Aromatic Hydrocarbons in TMC-1 CP and MC27/L1521F}

The data compiled in Fig.~\ref{fig:abundance} is listed in Table~\ref{tab:CDcomp}. The references these values originate from are given.

\begin{table*}[!htb]
    \centering
    \caption{Column densities of molecular hydrogen and carbon-bearing cycles observed in TMC-1 CP and MC27/L1521F as presented in Fig.~\ref{fig:abundance}.}    
    \begin{tabular}{c c c c c c}
    \toprule
\multicolumn{5}{c}{\textbf{Molecules in TMC-1 CP}} \\
    \midrule
    Molecule & Chemical Formula & $N_\mathrm{C}$ & Column Density & Reference \\
              &   &   &    ($10^{12}\,\mathrm{cm}^{-2}$)  \\
     \midrule
    Molecular hydrogen & \ce{H2} & 0 & 1.82$\times 10^{10}$ &~\citet{fuente2019,kirk2024} \\   
    Benzonitrile & \ce{C6H5CN} & 7 & 1.65$^{+0.09}_{-0.06}$ &~\citet{xue2025}\\
    1H-indene & \ce{C9H8}      & 9 &  11.4$^{+3.7}_{-2.2}$ &~\citet{xue2025}                \\
    2-cyanoindene & 2-\ce{C9H7CN}   & 10 &   0.15$^{+0.08}_{-0.03}$       &~\citet{xue2025}           \\
    1H-cyclopent[cd]indene & \ce{C11H8} & 11 &  6.0$^{+0.5}_{-0.5}$  &~\citet{fuentetaja2026}\\
    1-cyanonaphthalene & 1-\ce{C10H7CN} & 11 &  0.80$^{+0.81}_{-0.16}$  &~\citet{xue2025}\\
    2-cyanonaphthalene & 2-\ce{C10H7CN} & 11 &  0.53$^{+0.14}_{-0.05}$ &~\citet{xue2025} \\
    1-cyanoacenaphthylene & 1-\ce{C12H7CN} & 13 & 1.09$^{+0.40}_{-0.18}$ &~\citet{xue2025} \\
    3-cyanoacenaphthylene & 3-\ce{C12H7CN} & 13 & 0.7$^{+0.07}_{-0.09}$ &~\citet{cernicharo2026} \\
    4-cyanoacenaphthylene & 4-\ce{C12H7CN} & 13 & 0.5$^{+0.06}_{-0.09}$ &~\citet{cernicharo2026} \\
    5-cyanoacenaphthylene & 5-\ce{C12H7CN} & 13 & 0.77$^{+0.38}_{-0.12}$  &~\citet{xue2025} \\
    1H-phenalene & \ce{C13H10} & 13 & 15.3$^{+2.7}_{-1.4}$  & This work  \\
    1-cyanopyrene & 1-\ce{C16H9CN} & 17 & 0.91$^{+0.19}_{-0.08}$ &~\citet{xue2025}\\
    2-cyanopyrene & 2-\ce{C16H9CN} & 17 & 0.57$^{+0.18}_{-0.08}$ &~\citet{xue2025}\\
    4-cyanopyrene & 4-\ce{C16H9CN} & 17 & 0.92$^{+0.35}_{-0.14}$ &~\citet{xue2025}\\
    Cyanocoronene & \ce{C24H11CN} & 25 & 2.69$^{+0.26}_{-0.23}$ &~\citet{wenzel2025c}\\

\midrule

\multicolumn{5}{c}{\textbf{Molecules in MC27/L1521F}} \\
    \midrule
    Molecule & Chemical Formula & $N_\mathrm{C}$ & Column Density & Reference \\
              &   &   &    ($10^{12}\,\mathrm{cm}^{-2}$)  \\
     \midrule
    Molecular hydrogen & \ce{H2} & 0 & 1.68$\times 10^{10}$ &~\citet{chitsazzadeh2014} \\
    Benzonitrile & \ce{C6H5CN} & 7 & 0.366$^{+0.056}_{-0.058}$ & This work\\
    1H-phenalene & \ce{C13H10} & 13 & 14.7$^{+4.9}_{-3.5}$  & This work  \\

    \bottomrule
    \end{tabular}
    \label{tab:CDcomp}
\end{table*}

\section{Ab-initio Characterization of the Reaction of CN and Phenalene}
\label{sec:calcs}

\subsection{Potential energy surface calculations}

An ab-initio potential energy surface for the reaction of CN and phenalene was calculated. Initial geometries for the reactants, products and adducts were produced in Avogadro \citep{Hanwell_2012_17}. Optimizations and harmonic frequency analyses were performed with the density functional theory (DFT) $\omega$B97M \citep{Chai_2008_084106,Mardirossian_2016_214110,Lehtola_2018_1} hybrid functional and the D4 empirical dispersion correction \citep{Caldeweyher__2019_154122,Caldeweyher_2020_8499}, along with the minimally augmented triple zeta Karlsruhe basis set ma-def2-TZVPP \citep{Weigend_2005_3297,Zheng_2011_295, Weigend_2006_1057} with very tight convergence criteria for both SCF and optimizations and the DEFGRID3 integration grid. Harmonic zero point energies were scaled for anharmonicity by 0.9779 \citep{Kesharwani_2014_1701}. 
Initial geometries for transition states were obtained from relaxed (all but scan coordinate minimized) scans of the forming or breaking of bonds. Following optimization, transition states were evaluated for the presence of a single imaginary mode, and intrinsic reaction coordinate scans (IRCs) \citep{Ishida_1977_2153,Neese_2020_224108} verified they linked products and reactants. Single point energies for $\omega$B97M-D4/ma-def2-TZVPP stationary points were refined using explicitly correlated couple cluster with singles and doubles and perturbative triple excitations CCSD(T)-F12c \citep{Adler_2007_221106,Knizia_2009_054104,Rauhut_2009_054105} including a resolution of integral (RI) approximation on the F12 component \citep{Valeev_2004_190,Noga_2009_1,Adler_2007_221106,Liakos_2013_2653}, with the correlation consistent double zeta basis set refined for F12 approaches cc-pVDZ-F12 \citep{Dunning_1989_1007,Kendall_1992_6796,Davidson_1996_514} orbital basis set and the aug-cc-pVTZ/c \citep{Dunning_1989_1007,Kendall_1992_6796,Weigend_2002_3175,Hatig_2005_59} and cc-pVDZ-F12-CABS  \citep{Peterson_2008_084102,Noga_2009_1,Valeev_2004_190} auxiliary basis sets. All calculations were performed with ORCA 6.1.0 \citep{Neese_2000_93,Neese_2012_73,Neese_2023_381,Neese_2025_15}. 

\begin{figure}[!htb]
    \centering
    \includegraphics[width=\linewidth]{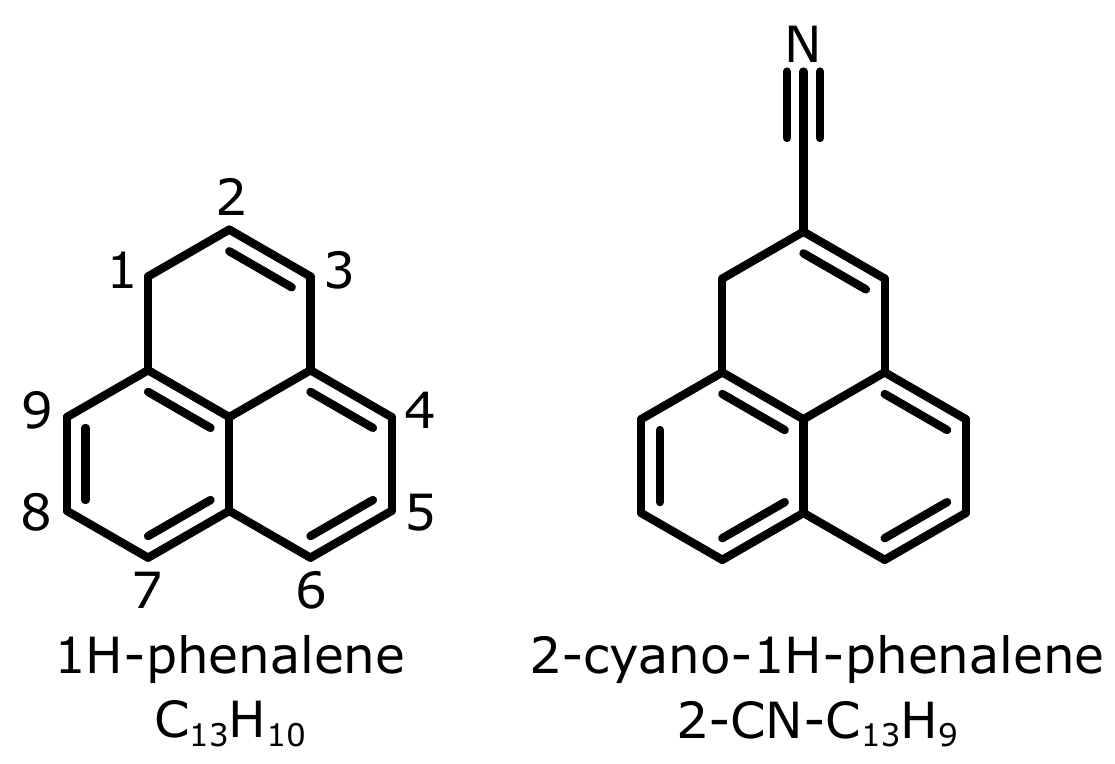}
    \caption{Structures of 1H-phenalene with its potential \textbf{substitution} sites and 2-cyano-1H-phenalene, as an example of one of the potential CN + phenalene products after H atom elimination.}
    \label{fig:cyanophenalene}
\end{figure}

\begin{figure*}
    \centering\includegraphics[width=\linewidth]{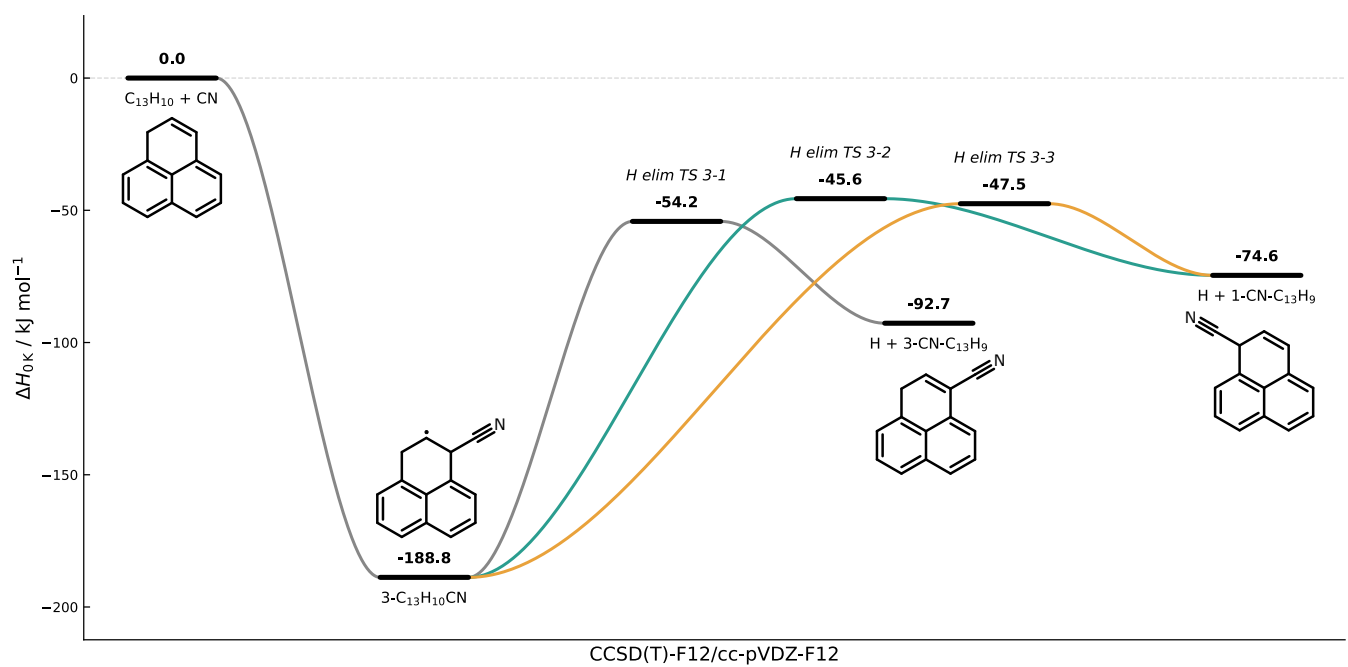}
    \caption{\textbf{A subset of the potential energy surface for the addition of CN to phenalene and subsequent H atom elimination channels at the CCSD(T)-F12/cc-pVDZ-F12//$\omega$B97M-D4/ma-def2-TZVPP level of theory. Only pathways upon addition to the 3-position (see Figure~\ref{fig:cyanophenalene}) are shown as an example. Other cyano-1H-phenalenes will form analogously and involving energies presented in Table~\ref{tab:CN_Phen_Energetics}.}}
    \label{fig:PES}
\end{figure*}

A complete characterization of the reaction of CN with phenalene is beyond the scope of this work due to the large number of sites. However, an analysis of the initial addition of CN to phenalene and the subsequent prompt elimination of an H atom from the carbon to which CN binds or from the CH$_2$ group was carried out. The addition of CN to the aromatic ring positions bearing a single H atom were all found to occur barrierlessly. The subsequent H atom elimination barriers, that result in a peri position $sp^3$ carbon, are found to be very submerged (-45--70\,kJ\,mol$^{-1}$). Therefore, formation of 2-, 3-, 4-, 5-, 6-, 7-, 8-cyano-1H-phenalene from their respective adducts is expected to be open at low temperatures. The barriers for H elimination yielding a non-peri $sp^3$ carbon were substantially emerged ($>$+85\,kJ\,mol$^{-1}$) with a very endothermic product (+160\,kJ\,mol$^{-1}$). As a result, 2-cyano-2H-phenalene is not predicted to form at low temperatures; whereas, 1-cyano-1H-phenalene can form following addition at the 3, 4, 6, 7, and 9 positions (see Figures~\ref{fig:cyanophenalene} and \ref{fig:PES} for example structures and potential energy surface, respectively).

\subsection{Prediction of site-specific rate coefficients}

The rate coefficient for the reactions of CN with phenalene and indene were estimated with classical capture theory \citep{Georgievskii_2005_194103,West_2019_134} in the manner applied in \citep{wenzel2025,wenzel2025c,Stewart_2025_11400,willis2026impact}. With ionization energies, polarizabilities and dipole moments from the Computational Chemistry Comparison and Benchmark DataBase \citep{CCCBDB-2022}, NIST chemistry webbook \citep{NISTwebbook}, or calculated with $\omega$B97M-D4/ma-def2-TZVPP.  At 10\,K, this predicts $k_{\textrm{CN+phen}} = 5.38^{+5.38}_{-2.69}\times 10^{-10}$\,cm$^3$s$^{-1}$ and $k_{\textrm{CN+ind}} = 4.88^{+4.88}_{-2.44} \times 10^{-10}$\,cm$^3$s$^{-1}$ for CN with phenalene and indene, respectively.

\begin{table*}[!htb]
    \centering
     \caption{Energetics of selected channels in the reaction of CN and phenalene. DFT refers to $\omega$B97M-D4/ma-def2-TZVPP, zpe scaled harmonic zero point energies, and CCSD(T)-F12 the CCSD(T)-F12/cc-pVDZ-F12 single point energies. $\Delta$H$_\mathrm{0K}$ are the relative energies including  ZPEs.}    
    \begin{tabular}{cccccc}
 \toprule
 & \multicolumn{3}{c}{Energies / Ha}& \multicolumn{2}{c}{$\Delta$H$_\mathrm{0K}$ /kJ\,mol$^{-1}$}\\
 Structures& DFT energy& zpe& CCSD(T)-F12& DFT&CCSD(T)-F12\\
  \midrule
         C$_{13}$H$_{10}$+CN&  -594.5320723&  0.190091178&  -593.1626747&  0.0& 0.0
\\
\midrule
         2-C$_{13}$H$_{10}$CN&  -594.6433633&  0.193911581&  -593.2628278&  -282.2& -252.9
\\
         H elim TS 2-1&  -594.5653908&  0.185834259&  
-593.1844863&  -98.7& 

-68.4\\
         H+2-CN-\ce{C13H9}&  -594.5718148&  0.184152848&  -593.1935108&  -119.9& -96.6
\\
 3-\ce{C13H10CN}& -594.6194126& 0.193201534& -593.2376985& -221.1&
-188.8\\
 H elim TS 3-1& -594.5599551& 0.185602329& -593.1788335& -85.0&
-54.2\\
 H+3-CN-\ce{C13H9}& -594.5703500& 0.18405476& -593.1919636& -116.3&-92.7
\\
         4-\ce{C13H10CN}&  -594.6220649&  0.192895701&  
-593.2414129&  -228.9& 
-199.4\\
 H elim TS 4-1& -594.5616031& 0.185630533& 
-593.1808713& -89.2&

-59.5\\
 H+4-CN-\ce{C13H9}& -594.5715500& 0.184017725& -593.1930133& -119.6&-95.6
\\
 5-\ce{C13H10CN}& -594.6128503& 0.192859458& 
-593.2338463& -204.8&
-179.6\\
 H elim TS 5-1& -594.5578435& 0.185546596& 
-593.1774470& -79.6&

-50.7\\
 H+5-CN-\ce{C13H9}& -594.5701196& 0.183924273& -593.1917687& -116.1&-92.6
\\
 6-\ce{C13H10CN}& -594.6298568& 0.193166406& 
-593.2496779& -248.7&
-220.4\\
 H elim TS 6-1& -594.5624101& 0.185623795& 
-593.1816438& -91.4&

-61.5\\
 H+6-CN-\ce{C13H9}& -594.5716307& 0.184062593& -593.193162& -119.7&-95.9
\\
 7-\ce{C13H10CN}& -594.6220803& 0.19287403& 
-593.2414155& -229.0&
-199.4\\
 H elim TS 7-1& -594.5612959& 0.185646406& 
-593.1801322& -88.4&

-57.5\\
 H+7-CN-\ce{C13H9}& -594.5714207& 0.18403922& -593.1929296& -119.2&-95.3
\\
 8-\ce{C13H10CN}& -594.6093772& 0.192575678& 
-593.2294700& -196.4&
-168.9\\
 H elim TS 8-1& -594.5572414& 0.185547271& 
-593.1763881& -78.0&

-47.9\\
 H+8-CN-\ce{C13H9}& -594.5701030& 0.183922023& -593.1917136& -116.0&-92.4
\\
 9-\ce{C13H10CN}& -594.6183946& 0.192914712& 
-593.2388733& -219.2&-192.6
\\
 H elim TS 9-1& -594.5625299& 0.185669984& 
-593.1815516& -91.6&

-61.2\\
 H+9-CN-\ce{C13H9}& -594.5726246& 0.183988152& -593.1940215& -122.5&-98.3
\\
\midrule
 \multicolumn{6}{c}{H elimination from CH$_2$ group in  adducts 3,4,6,7,9}\\
 \midrule
 H elim TS 3-2& -594.5565493& 0.185691587& 
-593.1756352& -75.8&
-45.6\\
 H elim TS 3-3& -594.5572444& 0.185578937& 
-593.1762520& -77.9&

-47.5\\
 H elim TS 4-2& -594.5574507& 0.185671372& -593.1766367& -78.2&
-48.3\\
 H elim TS 4-3& -594.557819& 0.185666238& 
-593.1769520& -79.2&
-49.1\\
 H elim TS 6-2& -594.558355& 0.185599024& 
-593.1778964& -80.8&

-51.8\\
 H elim TS 6-3& -594.5585252& 0.185597362& -593.1780215& -81.3&
-52.1\\
 H elim TS 7-2& -594.5578791& 0.185647129& 
-593.1774521& -79.4&
-50.5\\
 H elim TS 7-3& -594.558122& 0.185639032& 
-593.1776427& -80.1&

-51.0\\
 H elim TS 9-2& -594.5565115& 0.185727614& -593.17615881& -75.6&
-46.9\\
 H elim TS 9-3& -594.5586103& 0.185658865& 
-593.1780109& -81.3&
-51.9\\
 H + 1-CN-\ce{C13H9}& -594.5634837& 0.184260979& -593.1852394& -97.8&
-74.6
\\
    \bottomrule
    \end{tabular}
    \label{tab:CN_Phen_Energetics}
\end{table*}

In the absence of an in-depth calculation of the long-range partitioning of the reactant flux into the adducts, it is pragmatic to treat the site specificity statistically. This approach predicts the observed ratio of 1-, 2-, and 4-cyanopyrene \citep{wenzel2024,wenzel2025} in TMC-1 CP, and is within a factor of two for 1-, 3-, 4-, and 5-cyanoacenaphthylene \citep{cernicharo2024,cernicharo2026}. If the initial attack yields prompt elimination of the H bound to the same carbon (which occurs via the lowest barriers in Table~\ref{tab:CN_Phen_Energetics}), this results in the formation of 2-, 3-, 4-, 5-, 6-, 7-, 8-, and 9-cyano-1H-phenalene in equal abundances with a rate coefficient of $k_{2-9} = 6.73^{+6.73}_{-3.36} \times 10^{-11}$\,cm$^3$s$^{-1}$ at 10\,K. If all submerged H elimination channels are equally competitive, 1-cyano-1H-phenalene will be the dominant product with $k_1 = 2.24^{+2.24}_{-1.12} \times 10^{-10}$\,cm$^3$s$^{-1}$, 2-, 5-, 8-cyano-1H-phenalene formation is unchanged with $k_{2,5,8} = 6.73^{+6.73}_{-3.36} \times 10^{-11}$\,cm$^3$s$^{-1}$ and 3-, 4-, 6-, 7-, 9-cyano-1H-phenalene formation is suppressed with $k_{3,4,6,7,9} = 2.24^{+2.24}_{-1.12} \times 10^{-11}$\,cm$^3$s$^{-1}$. 

\subsection{Estimation of the abundance of different isomers of cyanophenalene in TMC-1}

The astrochemical modeling in \citet{wenzel2025} derived the ratio of cyano-tagged benzene and indene from both the chemical age and the CN + aromatic rate coefficient. Here, the ratios predicted at $2 \times 10^{5}$ years are used in combination with the rate coefficients above, to predict the abundances of the CN tagged species from their bare PAH parents with the results summarized in Table~\ref{tab:CNtoHRatios}. There is a reasonable agreement between the predicted and observed column densities of 2-cyanoindene in TMC-1 CP, if equal formation of each isomer occurs. Under the same approach 1/8th of the reaction flux results in the formation of each of 2-, 3-, 4-, 5-, 6-, 7-, 8-, and 9-cyano-1H-phenalene (9-CN-\ce{C13H9}) with an abundance in TMC-1 CP of $N(\ce{C13H9CN}) = 2.72^{+3.70}_{-1.49} \times 10^{11}\,\mathrm{cm^{-2}}$. However, if all submerged H elimination channels are equally competitive, this value could rise to $N(\ce{C13H9CN}) = 9.07^{+12.34}_{-4.98} \times 10^{11}\,\mathrm{cm^{-2}}$ for 1-cyano-1H-phenalene (1-CN-\ce{C13H9}).

\begin{table*}[htb]
    \centering
    \caption{Observed (obs) and predicted (pred) parent to child ratios (H/CN) for cyano functionalized indene and phenalene isomers in TMC-1 CP. Observed column densities are taken from recent observations of TMC-1 CP; from this work (1)  and from \citet{xue2025} (2). The site-specific scenario refers to 1/6th and 1/8th of flux per isomer, For CN + indene and CN + phenalene, respectively.}
    \begin{tabular}{lcc}
    \toprule
 & Phenalene (1)& Indene (2)\\
         \midrule
         $N$(H) (10$^{13}$\,cm$^{-2}) $&  1.53$^{+0.27}_{-0.14}$&  1.14$^{+0.37}_{-0.22}$\\
        $N$(CN)$_{\textrm{obs}}$ (10$^{11}$\,cm$^{-2})$&  --- &  
1.50$^{+0.80}_{-0.30}$\\
         $[\mathrm{H/CN}]_{\textrm{obs}}$&  ---&  76$^{+49}_{-36}$\\
         Total $[\mathrm{H/CN}]_{\textrm{pred}}$&  7.03$^{+7.12}_{-3.52}$&  7.75$^{+7.86}_{-3.89}$\\
        Site specific $[\mathrm{H/CN}]_{\textrm{pred}}$&  
56.2$^{+57.0}_{-28.2}$&  46.5$^{+47.2}_{-23.3}$\\
 Total $N$(CN)$_{\textrm{pred}}$ (10$^{12}$\,cm$^{-2}$) & 2.18$^{+2.96}_{-1.19}$& 1.47$^{+2.44}_{-0.88}$\\
 Site specific $N$(CN)$_{\textrm{pred}}$  (10$^{11}$\,cm$^{-2}$) & 2.72$^{+3.70}_{-1.49}$& 2.45$^{+4.07}_{-1.47}$\\
 \bottomrule
    \end{tabular}
    \label{tab:CNtoHRatios}
\end{table*}

\bibliography{references, calcs}{}

@article{Weigend_2005_3297,
	title = {Balanced basis sets of split valence, triple zeta valence and quadruple zeta valence quality for {H} to {Rn}: {Design} and assessment of accuracy},
	volume = {7},
	issn = {1463-9084},
	shorttitle = {Balanced basis sets of split valence, triple zeta valence and quadruple zeta valence quality for {H} to {Rn}},
	url = {https://pubs.rsc.org/en/content/articlelanding/2005/cp/b508541a},
	doi = {10.1039/B508541A},
	language = {en},
	number = {18},
	urldate = {2026-01-19},
	journal = {Physical Chemistry Chemical Physics},
	publisher = {The Royal Society of Chemistry},
	author = {Weigend, Florian and Ahlrichs, Reinhart},
	month = aug,
	year = {2005},
	pages = {3297--3305},
}

@article{Zheng_2011_295,
	title = {Minimally augmented {Karlsruhe} basis sets},
	volume = {128},
	issn = {1432-2234},
	url = {https://doi.org/10.1007/s00214-010-0846-z},
	doi = {10.1007/s00214-010-0846-z},
	language = {en},
	number = {3},
	urldate = {2026-01-19},
	journal = {Theoretical Chemistry Accounts},
	author = {Zheng, Jingjing and Xu, Xuefei and Truhlar, Donald G.},
	month = feb,
	year = {2011},
	pages = {295--305},
}

@article{Weigend_2006_1057,
	title = {Accurate {Coulomb}-fitting basis sets for {H} to {Rn}},
	volume = {8},
	issn = {1463-9084},
	url = {https://pubs.rsc.org/en/content/articlelanding/2006/cp/b515623h},
	doi = {10.1039/B515623H},
	language = {en},
	number = {9},
	urldate = {2026-01-19},
	journal = {Physical Chemistry Chemical Physics},
	publisher = {The Royal Society of Chemistry},
	author = {Weigend, Florian},
	month = feb,
	year = {2006},
	pages = {1057--1065},
}

@article{Ishida_1977_2153,
	title = {The intrinsic reaction coordinate. {An} ab initio calculation for {HNC}→{HCN} and {H}−+{CH4}→{CH4}+{H}−},
	volume = {66},
	issn = {0021-9606},
	url = {https://doi.org/10.1063/1.434152},
	doi = {10.1063/1.434152},
	number = {5},
	urldate = {2026-01-19},
	journal = {The Journal of Chemical Physics},
	author = {Ishida, Kazuhiro and Morokuma, Keiji and Komornicki, Andrew},
	month = mar,
	year = {1977},
	pages = {2153--2156},
}

@article{Neese_2020_224108,
  author = {Neese, Frank and Wennmohs, Frank and Becker, Ute and Riplinger, Christoph},
  title = {The ORCA Quantum Chemistry Program Package},
  journal = {The Journal of Chemical Physics},
  year = {2020},
  volume = {152},
  number = {22},
  pages = {224108},
  doi = {10.1063/5.0004608}
}

@article{Kesharwani_2014_1701,
	title = {Frequency and {Zero}-{Point} {Vibrational} {Energy} {Scale} {Factors} for {Double}-{Hybrid} {Density} {Functionals} (and {Other} {Selected} {Methods}): {Can} {Anharmonic} {Force} {Fields} {Be} {Avoided}?},
	volume = {119},
	issn = {1089-5639},
	shorttitle = {Frequency and {Zero}-{Point} {Vibrational} {Energy} {Scale} {Factors} for {Double}-{Hybrid} {Density} {Functionals} (and {Other} {Selected} {Methods})},
	url = {https://doi.org/10.1021/jp508422u},
	doi = {10.1021/jp508422u},
	number = {9},
	urldate = {2026-01-19},
	journal = {The Journal of Physical Chemistry A},
	publisher = {American Chemical Society},
	author = {Kesharwani, Manoj K. and Brauer, Brina and Martin, Jan M. L.},
	month = mar,
	year = {2014},
	pages = {1701--1714},
}

@article{Caldeweyher_2020_8499,
	title = {Extension and evaluation of the {D4} {London}-dispersion model for periodic systems},
	volume = {22},
	issn = {1463-9084},
	url = {https://pubs.rsc.org/en/content/articlelanding/2020/cp/d0cp00502a},
	doi = {10.1039/D0CP00502A},
	language = {en},
	number = {16},
	urldate = {2026-01-19},
	journal = {Physical Chemistry Chemical Physics},
	publisher = {The Royal Society of Chemistry},
	author = {Caldeweyher, Eike and Mewes, Jan-Michael and Ehlert, Sebastian and Grimme, Stefan},
	month = apr,
	year = {2020},
	pages = {8499--8512},
}

@article{Caldeweyher__2019_154122,
	title = {A generally applicable atomic-charge dependent {London} dispersion correction},
	volume = {150},
	issn = {0021-9606},
	url = {https://doi.org/10.1063/1.5090222},
	doi = {10.1063/1.5090222},
	number = {15},
	urldate = {2026-01-19},
	journal = {The Journal of Chemical Physics},
	author = {Caldeweyher, Eike and Ehlert, Sebastian and Hansen, Andreas and Neugebauer, Hagen and Spicher, Sebastian and Bannwarth, Christoph and Grimme, Stefan},
	month = apr,
	year = {2019},
	pages = {154122},
}

@article{Chai_2008_084106,
	title = {Systematic optimization of long-range corrected hybrid density functionals},
	volume = {128},
	issn = {0021-9606},
	url = {https://doi.org/10.1063/1.2834918},
	doi = {10.1063/1.2834918},
	number = {8},
	urldate = {2026-01-19},
	journal = {The Journal of Chemical Physics},
	author = {Chai, Jeng-Da and Head-Gordon, Martin},
	month = feb,
	year = {2008},
	pages = {084106},
}

@article{Mardirossian_2016_214110,
	title = {ω{B97M}-{V}: {A} combinatorially optimized, range-separated hybrid, meta-{GGA} density functional with {VV10} nonlocal correlation},
	volume = {144},
	issn = {0021-9606},
	shorttitle = {ω{B97M}-{V}},
	url = {https://doi.org/10.1063/1.4952647},
	doi = {10.1063/1.4952647},
	number = {21},
	urldate = {2026-01-19},
	journal = {The Journal of Chemical Physics},
	author = {Mardirossian, Narbe and Head-Gordon, Martin},
	month = jun,
	year = {2016},
	pages = {214110},
}

@article{Lehtola_2018_1,
	title = {Recent developments in libxc — {A} comprehensive library of functionals for density functional theory},
	volume = {7},
	issn = {2352-7110},
	url = {https://www.sciencedirect.com/science/article/pii/S2352711017300602},
	doi = {10.1016/j.softx.2017.11.002},
	urldate = {2026-01-19},
	journal = {SoftwareX},
	author = {Lehtola, Susi and Steigemann, Conrad and Oliveira, Micael J. T. and Marques, Miguel A. L.},
	month = jan,
	year = {2018},
	pages = {1--5},
}

@article{Hanwell_2012_17,
  author = {Hanwell, Marcus D. and Curtis, Donald E. and Lonie, David C. and Vandermeersch, Tim and Zurek, Eva and Hutchison, Geoffrey R.},
  title = {Avogadro: an advanced semantic chemical editor, visualization, and analysis platform},
  journal = {Journal of Cheminformatics},
  year = {2012},
  volume = {4},
  number = {1},
  pages = {17},
  doi = {10.1186/1758-2946-4-17},
  url = {https://doi.org/10.1186/1758-2946-4-17},
  issn = {1758-2946},
}

@article{Neese_2025_15,
	title = {Software {Update}: {The} {ORCA} {Program} {System}—{Version} 6.0},
	volume = {15},
	shorttitle = {Software {Update}},
	doi = {10.1002/wcms.70019},
	journal = {WIREs Computational Molecular Science},
	author = {Neese, Frank},
	month = apr,
	year = {2025},
}

@article{Neese_2023_381,
	title = {The {SHARK} integral generation and digestion system},
	volume = {44},
	copyright = {© 2022 The Author. Journal of Computational Chemistry published by Wiley Periodicals LLC.},
	issn = {1096-987X},
	url = {https://onlinelibrary.wiley.com/doi/abs/10.1002/jcc.26942},
	doi = {10.1002/jcc.26942},
	language = {en},
	number = {3},
	urldate = {2026-01-19},
	journal = {Journal of Computational Chemistry},
	author = {Neese, Frank},
	year = {2023},
	note = {\_eprint: https://onlinelibrary.wiley.com/doi/pdf/10.1002/jcc.26942},
	pages = {381--396},
}

@article{Neese_2012_73,
	title = {The {ORCA} program system},
	volume = {2},
	copyright = {Copyright © 2011 John Wiley \& Sons, Ltd.},
	issn = {1759-0884},
	url = {https://onlinelibrary.wiley.com/doi/abs/10.1002/wcms.81},
	doi = {10.1002/wcms.81},
	language = {en},
	number = {1},
	urldate = {2026-01-19},
	journal = {WIREs Computational Molecular Science},
	author = {Neese, Frank},
	year = {2012},
	note = {\_eprint: https://wires.onlinelibrary.wiley.com/doi/pdf/10.1002/wcms.81},
	pages = {73--78},
}

@article{Neese_2000_93,
	title = {Approximate second-order {SCF} convergence for spin unrestricted wavefunctions},
	volume = {325},
	issn = {0009-2614},
	url = {https://www.sciencedirect.com/science/article/pii/S000926140000662X},
	doi = {10.1016/S0009-2614(00)00662-X},
	number = {1},
	urldate = {2026-01-19},
	journal = {Chemical Physics Letters},
	author = {Neese, Frank},
	month = jul,
	year = {2000},
	pages = {93--98},
}

@article{Dunning_1989_1007,
  author = {Dunning, Jr., Thom H.},
  title = {Gaussian Basis Sets for Use in Correlated Molecular Calculations. I. The Atoms Boron through Neon and Hydrogen},
  journal = {The Journal of Chemical Physics},
  year = {1989},
  volume = {90},
  number = {2},
  pages = {1007--1023},
  doi = {10.1063/1.456153}
}

@article{Kendall_1992_6796,
  author = {Kendall, Rick A. and Dunning, Jr., Thom H. and Harrison, Robert J.},
  title = {Electron Affinities of the First-Row Atoms Revisited. Systematic Basis Sets and Wave Functions},
  journal = {The Journal of Chemical Physics},
  year = {1992},
  volume = {96},
  number = {9},
  pages = {6796--6806},
  doi = {10.1063/1.462569}
}

@article{Weigend_2002_3175,
  author = {Weigend, Florian and Kohn, Andreas and Hattig, Christof},
  title = {Efficient Use of the Correlation Consistent Basis Sets in Resolution of the Identity MP2 Calculations},
  journal = {The Journal of Chemical Physics},
  year = {2002},
  volume = {116},
  number = {8},
  pages = {3175--3183},
  doi = {10.1063/1.1445115}
}

@article{Hatig_2005_59,
  author = {Hattig, Christof},
  title = {Optimization of Auxiliary Basis Sets for RI-MP2 and RI-CC2 Calculations: Core-Valence and Quintuple-Zeta Basis Sets for H to Ar and QZVPP Basis Sets for Li to Kr},
  journal = {Physical Chemistry Chemical Physics},
  year = {2005},
  volume = {7},
  number = {1},
  pages = {59--66},
  doi = {10.1039/B415208E}
}

@article{Davidson_1996_514,
  author = {Davidson, Ernest R.},
  title = {Comment on ``Comment on Dunning's Correlation-Consistent Basis Sets''},
  journal = {Chemical Physics Letters},
  year = {1996},
  volume = {260},
  number = {3--4},
  pages = {514--518},
  doi = {10.1016/0009-2614(96)00917-7}
}

@article{Peterson_2008_084102,
  author = {Peterson, Kirk A. and Adler, Thomas B. and Werner, Hans-Joachim},
  title = {Systematically Convergent Basis Sets for Explicitly Correlated Wavefunctions: The Atoms H, He, B-Ne, and Al-Ar},
  journal = {The Journal of Chemical Physics},
  year = {2008},
  volume = {128},
  number = {8},
  pages = {084102},
  doi = {10.1063/1.2831537}
}

@article{Noga_2009_1,
  author = {Noga, Jozef and Simunek, Jiri},
  title = {On the One-Particle Basis Set Relaxation in R12-Based Theories},
  journal = {Chemical Physics},
  year = {2009},
  volume = {356},
  number = {1--3},
  pages = {1--6},
  doi = {10.1016/j.chemphys.2008.10.012}
}

@article{Valeev_2004_190,
  author = {Valeev, Edward F.},
  title = {Improving on the Resolution of the Identity in Linear R12 Ab Initio Theories},
  journal = {Chemical Physics Letters},
  year = {2004},
  volume = {395},
  number = {4--6},
  pages = {190--195},
  doi = {10.1016/j.cplett.2004.07.061}
}

@article{Adler_2007_221106,
    author = {Adler, Thomas B. and Knizia, Gerald and Werner, Hans-Joachim},
    title = {A simple and efficient CCSD(T)-F12 approximation},
    journal = {The Journal of Chemical Physics},
    volume = {127},
    number = {22},
    pages = {221106},
    year = {2007},
    month = {12},
    issn = {0021-9606},
    doi = {10.1063/1.2817618},
    url = {https://doi.org/10.1063/1.2817618},
    eprint = {https://pubs.aip.org/aip/jcp/article-pdf/doi/10.1063/1.2817618/13499829/221106\_1\_online.pdf},
}

@article{Knizia_2009_054104,
    author = {Knizia, Gerald and Adler, Thomas B. and Werner, Hans-Joachim},
    title = {Simplified CCSD(T)-F12 methods: Theory and benchmarks},
    journal = {The Journal of Chemical Physics},
    volume = {130},
    number = {5},
    pages = {054104},
    year = {2009},
    month = {02},
    issn = {0021-9606},
    doi = {10.1063/1.3054300},
    url = {https://doi.org/10.1063/1.3054300},
    eprint = {https://pubs.aip.org/aip/jcp/article-pdf/doi/10.1063/1.3054300/15424674/054104\_1\_online.pdf},
}

@article{Rauhut_2009_054105,
    author = {Rauhut, Guntram and Knizia, Gerald and Werner, Hans-Joachim},
    title = {Accurate calculation of vibrational frequencies using explicitly correlated coupled-cluster theory},
    journal = {The Journal of Chemical Physics},
    volume = {130},
    number = {5},
    pages = {054105},
    year = {2009},
    month = {02},
    issn = {0021-9606},
    doi = {10.1063/1.3070236},
    url = {https://doi.org/10.1063/1.3070236},
    eprint = {https://pubs.aip.org/aip/jcp/article-pdf/doi/10.1063/1.3070236/15422793/054105\_1\_online.pdf},
}

@article{Liakos_2013_2653,
author = {Dimitrios G. Liakos and Róbert Izsák and Edward F. Valeev and Frank Neese},
title = {What is the most efficient way to reach the canonical MP2 basis set limit?},
journal = {Molecular Physics},
volume = {111},
number = {16-17},
pages = {2653--2662},
year = {2013},
publisher = {Taylor \& Francis},
doi = {10.1080/00268976.2013.824624},
URL = { 
        https://doi.org/10.1080/00268976.2013.824624
},
eprint = {    
        https://doi.org/10.1080/00268976.2013.824624
}
}

@article{Georgievskii_2005_194103,
  title = {Long-Range Transition State Theory},
  author = {Georgievskii, Yuri and Klippenstein, Stephen J.},
  year = {2005},
  month = may,
  journal = {The Journal of Chemical Physics},
  volume = {122},
  number = {19},
  pages = {194103},
  issn = {0021-9606},
  doi = {10.1063/1.1899603},
  urldate = {2024-06-14}
}

@article{West_2019_134,
  title = {Measurements of {{Low Temperature Rate Coefficients}} for the {{Reaction}} of {{CH}} with {{CH2O}} and {{Application}} to {{Dark Cloud}} and {{AGB Stellar Wind Models}}},
  author = {West, Niclas A. and Millar, Tom J. and de Sande, Marie Van and Rutter, Edward and Blitz, Mark A. and Decin, Leen and Heard, Dwayne E.},
  year = {2019},
  month = nov,
  journal = {The Astrophysical Journal},
  volume = {885},
  number = {2},
  pages = {134},
  publisher = {The American Astronomical Society},
  issn = {0004-637X},
  doi = {10.3847/1538-4357/ab480e},
  urldate = {2024-06-14},
  langid = {english}
}

@article{Stewart_2025_11400,
author = {Stewart, D. Archie and Speak, Thomas H. and Yuan, Elsa Q. H. and Willis, Reace H. J. and Sconce, Victoria B. and Fried, Zachary T. P. and Holdren, Martin S. and Lipnicky, Andrew and Shingledecker, Christopher and McCarthy, Michael C. and Cooke, Ilsa R. and Wenzel, Gabi and McGuire, Brett A.},
title = {Centimeter-Wave Rotational Spectroscopy of Ethynylbenzonitriles: Structural Analysis and Astronomical Search},
journal = {The Journal of Physical Chemistry A},
volume = {129},
number = {49},
pages = {11400-11413},
year = {2025},
doi = {10.1021/acs.jpca.5c05772},
    note ={PMID: 41326172},
URL = { 
        https://doi.org/10.1021/acs.jpca.5c05772
},
eprint = { 
        https://doi.org/10.1021/acs.jpca.5c05772
}
}

@article{stockett_efficient_2023,
    title = {Efficient stabilization of cyanonaphthalene by fast radiative cooling and implications for the resilience of small {PAHs} in interstellar clouds},
    volume = {14},
    copyright = {2023 The Author(s)},
    issn = {2041-1723},
    url = {https://www.nature.com/articles/s41467-023-36092-0},
    doi = {10.1038/s41467-023-36092-0},
    language = {en},
    number = {1},
    urldate = {2024-07-08},
    journal = {Nature Communications},
    author = {Stockett, Mark H. and Bull, James N. and Cederquist, Henrik and Indrajith, Suvasthika and Ji, MingChao and Navarro Navarrete, José E. and Schmidt, Henning T. and Zettergren, Henning and Zhu, Boxing},
    month = jan,
    year = {2023},
    pages = {395},
}

@article{johansson2018,
    title = {Resonance-stabilized hydrocarbon-radical chain reactions may explain soot inception and growth},
    volume = {361},
    url = {https://www.science.org/doi/10.1126/science.aat3417},
    doi = {10.1126/science.aat3417},
    number = {6406},
    urldate = {2026-05-15},
    journal = {Science},
    author = {Johansson, K. O. and Head-Gordon, M. P. and Schrader, P. E. and Wilson, K. R. and Michelsen, H. A.},
    month = sep,
    year = {2018},
    note = {Publisher: American Association for the Advancement of Science},
    pages = {997--1000},
}

@article{willis2026impact,
  title={The impact of hydrogen atom tunneling on aromatic chemistry in TMC-1},
  author={Willis, Reace H J and Speak, Thomas H and Byrne, Alex N and Shingledecker, Christopher N and Cooke, Ilsa R},
  journal={arXiv preprint arXiv:2604.21892},
  year={2026}
}

@article{oconnor2017,
	title = {Hydrogen-adduction to open-shell graphene fragments: spectroscopy, thermochemistry and astrochemistry},
	volume = {8},
	issn = {2041-6539},
	shorttitle = {Hydrogen-adduction to open-shell graphene fragments},
	url = {https://pubs.rsc.org/en/content/articlelanding/2017/sc/c6sc03787a},
	doi = {10.1039/C6SC03787A},
	language = {en},
	number = {2},
	urldate = {2026-06-21},
	journal = {Chemical Science},
	publisher = {The Royal Society of Chemistry},
	author = {O'Connor, Gerard D. and Chan, Bun and Sanelli, Julian A. and Cergol, Katie M. and Dryza, Viktoras and Payne, Richard J. and Bieske, Evan J. and Radom, Leo and Schmidt, Timothy W.},
	month = jan,
	year = {2017},
	pages = {1186--1194},
}

@article{oconnor2011,
	title = {Spectroscopy of the {Free} {Phenalenyl} {Radical}},
	volume = {133},
	issn = {0002-7863},
	url = {https://doi.org/10.1021/ja206322n},
	doi = {10.1021/ja206322n},
	number = {37},
	urldate = {2026-06-21},
	journal = {Journal of the American Chemical Society},
	publisher = {American Chemical Society},
	author = {O’Connor, Gerard D. and Troy, Tyler P. and Roberts, Derrick A. and Chalyavi, Nahid and Fückel, Burkhard and Crossley, Maxwell J. and Nauta, Klaas and Stanton, John F. and Schmidt, Timothy W.},
	month = sep,
	year = {2011},
	pages = {14554--14557},
}

@article{crapsi2004,
	title = {Observations of {L1521F}: {A} highly evolved starless core},
	volume = {420},
	copyright = {© ESO, 2004},
	issn = {0004-6361, 1432-0746},
	shorttitle = {Observations of {L1521F}},
	url = {https://www.aanda.org/articles/aa/abs/2004/24/aa0915/aa0915.html},
	doi = {10.1051/0004-6361:20035915},
	language = {en},
	number = {3},
	urldate = {2026-05-25},
	journal = {Astronomy \& Astrophysics},
	publisher = {EDP Sciences},
	author = {Crapsi, A. and Caselli, P. and Walmsley, C. M. and Tafalla, M. and Lee, C. W. and Bourke, T. L. and Myers, P. C.},
	month = jun,
	year = {2004},
	pages = {957--974},
}

@article{bull2025,
	title = {Radiative {Stabilization} of the {Indenyl} {Cation}: {Recurrent} {Fluorescence} in a {Closed}-{Shell} {Polycyclic} {Aromatic} {Hydrocarbon}},
	volume = {134},
	shorttitle = {Radiative {Stabilization} of the {Indenyl} {Cation}},
	url = {https://link.aps.org/doi/10.1103/PhysRevLett.134.228002},
	doi = {10.1103/PhysRevLett.134.228002},
	number = {22},
	urldate = {2026-05-25},
	journal = {Physical Review Letters},
	publisher = {American Physical Society},
	author = {Bull, James N. and Subramani, Arun and Liu, Chang and Marlton, Samuel J. P. and Ashworth, Eleanor K. and Cederquist, Henrik and Zettergren, Henning and Stockett, Mark H.},
	month = jun,
	year = {2025},
	pages = {228002},
}

@article{kamer2023,
	title = {Threshold photoelectron spectroscopy and dissociative photoionization of benzonitrile},
	volume = {25},
	issn = {1463-9084},
	url = {https://pubs.rsc.org/en/content/articlelanding/2023/cp/d3cp03977c},
	doi = {10.1039/D3CP03977C},
	language = {en},
	number = {42},
	urldate = {2026-05-25},
	journal = {Physical Chemistry Chemical Physics},
	publisher = {The Royal Society of Chemistry},
	author = {Kamer, Jerry and Schleier, Domenik and Donker, Merel and Hemberger, Patrick and Bodi, Andras and Bouwman, Jordy},
	month = nov,
	year = {2023},
	pages = {29070--29079},
}

@article{debes2025,
	title = {Sequential dissociation of ionized benzonitrile: {New} pathways to reactive interstellar ions and neutrals},
	volume = {693},
	copyright = {© The Authors 2025},
	issn = {0004-6361, 1432-0746},
	shorttitle = {Sequential dissociation of ionized benzonitrile},
	url = {https://www.aanda.org/articles/aa/abs/2025/01/aa49818-24/aa49818-24.html},
	doi = {10.1051/0004-6361/202449818},
	language = {en},
	urldate = {2026-05-25},
	journal = {Astronomy \& Astrophysics},
	publisher = {EDP Sciences},
	author = {Debes, D. Bou and Mendes, M. and Rodrigues, R. and Ameixa, J. and Cornetta, L. M. and Silva, F. Ferreira da and Eden, S.},
	month = jan,
	year = {2025},
	pages = {A304},
}

@article{bergin2003,
	title = {The {Effects} of {UV} {Continuum} and {Lyα} {Radiation} on the {Chemical} {Equilibrium} of {T} {Tauri} {Disks}},
	volume = {591},
	issn = {0004-637X},
	url = {https://iopscience.iop.org/article/10.1086/377148},
	doi = {10.1086/377148},
	language = {en},
	number = {2},
	urldate = {2026-05-24},
	journal = {The Astrophysical Journal},
	publisher = {IOP Publishing},
	author = {Bergin, Edwin and Calvet, Nuria and D’Alessio, Paola and Herczeg, Gregory J.},
	month = jun,
	year = {2003},
	pages = {L159},
}

@article{yang2011,
	title = {A {FAR}-{ULTRAVIOLET} {ATLAS} {OF} {LOW}-{RESOLUTION} {HUBBLE} {SPACE} ℡{ESCOPE} {SPECTRA} {OF} {T} {TAURI} {STARS}*},
	volume = {744},
	issn = {0004-637X},
	url = {https://doi.org/10.1088/0004-637X/744/2/121},
	doi = {10.1088/0004-637X/744/2/121},
	language = {en},
	number = {2},
	urldate = {2026-05-24},
	journal = {The Astrophysical Journal},
	publisher = {The American Astronomical Society},
	author = {Yang, Hao and Herczeg, Gregory J. and Linsky, Jeffrey L. and Brown, Alexander and Johns-Krull, Christopher M. and Ingleby, Laura and Calvet, Nuria and Bergin, Edwin and Valenti, Jeff A.},
	month = dec,
	year = {2011},
	pages = {121},
}

@article{fuente2019,
	title = {Gas phase {Elemental} abundances in {Molecular} {cloudS} ({GEMS}) - {I}. {The} prototypical dark cloud {TMC} 1},
	volume = {624},
	copyright = {© ESO 2019},
	issn = {0004-6361, 1432-0746},
	url = {https://www.aanda.org/articles/aa/abs/2019/04/aa34654-18/aa34654-18.html},
	doi = {10.1051/0004-6361/201834654},
	language = {en},
	urldate = {2024-05-06},
	journal = {Astronomy \& Astrophysics},
	publisher = {EDP Sciences},
	author = {Fuente, A. and Navarro, D. G. and Caselli, P. and Gerin, M. and Kramer, C. and Roueff, E. and Alonso-Albi, T. and Bachiller, R. and Cazaux, S. and Commercon, B. and Friesen, R. and García-Burillo, S. and Giuliano, B. M. and Goicoechea, J. R. and Gratier, P. and Hacar, A. and Jiménez-Serra, I. and Kirk, J. and Lattanzi, V. and Loison, J. C. and Malinen, J. and Marcelino, N. and Martín-Doménech, R. and Muñoz-Caro, G. and Pineda, J. and Tafalla, M. and Tercero, B. and Ward-Thompson, D. and Treviño-Morales, S. P. and Riviére-Marichalar, P. and Roncero, O. and Vidal, T. and Ballester, M. Y.},
	month = apr,
	year = {2019},
	pages = {A105},
}

@article{kirk2024,
	title = {Herschel {Gould} {Belt} {Survey} in {Taurus} – {II}. {A} census of dense cores and filaments in the {TMC1} region},
	volume = {532},
	issn = {0035-8711},
	url = {https://doi.org/10.1093/mnras/stae1633},
	doi = {10.1093/mnras/stae1633},
	number = {4},
	urldate = {2026-05-22},
	journal = {Monthly Notices of the Royal Astronomical Society},
	author = {Kirk, J M and Ward-Thompson, D and Di Francesco, J and André, Ph and Bresnahan, D W and Könyves, V and Marsh, K and Griffin, M J and Schneider, N and Men’shchikov, A and Palmeirim, P and Bontemps, S and Arzoumanian, D and Benedettini, M and Pezzuto, S},
	month = aug,
	year = {2024},
	pages = {4661--4680},
}

@misc{scholler2026,
	title = {Tracing the sulfur depletion in starless and pre-stellar cores},
	url = {http://arxiv.org/abs/2605.13635},
	doi = {10.48550/arXiv.2605.13635},
	urldate = {2026-05-22},
	publisher = {arXiv},
	author = {Schöller, L. and Spezzano, S. and Sipilä, O. and Makarenko, E. I. and Caselli, P. and Bunn, H. A. and Jensen, S. S.},
	month = may,
	year = {2026},
	note = {arXiv:2605.13635 [astro-ph.GA]},
}

@article{favre2020,
	title = {Seeds of {Life} in {Space} ({SOLIS}) - {VII}. {Discovery} of a cold dense methanol blob toward the {L1521F} {VeLLO} system},
	volume = {635},
	copyright = {© C. Favre et al. 2020},
	issn = {0004-6361, 1432-0746},
	url = {https://www.aanda.org/articles/aa/abs/2020/03/aa37297-19/aa37297-19.html},
	doi = {10.1051/0004-6361/201937297},
	language = {en},
	urldate = {2026-01-27},
	journal = {Astronomy \& Astrophysics},
	publisher = {EDP Sciences},
	author = {Favre, C. and Vastel, C. and Jimenez-Serra, I. and Quénard, D. and Caselli, P. and Ceccarelli, C. and Chacón-Tanarro, A. and Fontani, F. and Holdship, J. and Oya, Y. and Punanova, A. and Sakai, N. and Spezzano, S. and Yamamoto, S. and Neri, R. and López-Sepulcre, A. and Alves, F. and Bachiller, R. and Balucani, N. and Bianchi, E. and Bizzocchi, L. and Codella, C. and Caux, E. and Simone, M. De and Romero, J. Enrique and Dulieu, F. and Feng, S. and Al-Edhari, A. Jaber and Lefloch, B. and Ospina-Zamudio, J. and Pineda, J. and Podio, L. and Rimola, A. and Segura-Cox, D. and Sims, I. R. and Taquet, V. and Testi, L. and Theulé, P. and Ugliengo, P. and Vasyunin, A. I. and Vazart, F. and Viti, S. and Witzel, A.},
	month = mar,
	year = {2020},
	pages = {A189},
}

@misc{gotham-spectral-pipeline,
	title = {{GOTHAM} spectral pipeline},
	url = {https://doi.org/10.5281/zenodo.15678187},
	doi = {10.5281/zenodo.15678187},
	publisher = {Zenodo},
	author = {Xue, Ci},
	month = jun,
	year = {2025},
}

@article{frenklach1985,
	series = {Twentieth {Symposium} ({International}) on {Combustion}},
	title = {Detailed kinetic modeling of soot formation in shock-tube pyrolysis of acetylene},
	volume = {20},
	issn = {0082-0784},
	url = {https://www.sciencedirect.com/science/article/pii/S0082078485805786},
	doi = {10.1016/S0082-0784(85)80578-6},
	number = {1},
	urldate = {2024-05-02},
	journal = {Symposium (International) on Combustion},
	author = {Frenklach, Michael and Clary, David W. and Gardiner, William C. and Stein, Stephen E.},
	month = jan,
	year = {1985},
	pages = {887--901},
}

@article{porfiriev2020,
	title = {Conversion of acenaphthalene to phenalene via methylation: {A} theoretical study},
	volume = {213},
	issn = {0010-2180},
	shorttitle = {Conversion of acenaphthalene to phenalene via methylation},
	url = {https://www.sciencedirect.com/science/article/pii/S0010218019305437},
	doi = {10.1016/j.combustflame.2019.11.038},
	urldate = {2026-02-11},
	journal = {Combustion and Flame},
	author = {Porfiriev, Denis P. and Azyazov, Valeriy N. and Mebel, Alexander M.},
	month = mar,
	year = {2020},
	pages = {302--313},
}

@article{loomis2016,
	title = {Non-detection of {HC11N} towards {TMC}-1: constraining the chemistry of large carbon-chain molecules},
	volume = {463},
	issn = {0035-8711},
	shorttitle = {Non-detection of {HC11N} towards {TMC}-1},
	url = {https://doi.org/10.1093/mnras/stw2302},
	doi = {10.1093/mnras/stw2302},
	number = {4},
	urldate = {2026-05-11},
	journal = {Monthly Notices of the Royal Astronomical Society},
	author = {Loomis, Ryan A. and Shingledecker, Christopher N. and Langston, Glen and McGuire, Brett A. and Dollhopf, Niklaus M. and Burkhardt, Andrew M. and Corby, Joanna and Booth, Shawn T. and Carroll, P. Brandon and Turner, Barry and Remijan, Anthony J.},
	month = dec,
	year = {2016},
	pages = {4175--4183},
}

@article{https://doi.org/10.1002/asna.200510446,
	title = {Spitzer discovery of very low luminosity objects},
	volume = {326},
	url = {https://onlinelibrary.wiley.com/doi/abs/10.1002/asna.200510446},
	doi = {https://doi.org/10.1002/asna.200510446},
	number = {10},
	journal = {Astronomische Nachrichten},
	author = {Kauffmann, J. and Bertoldi, F. and Evans II, N. J. and c2d Collaboration, The},
	year = {2005},
	note = {tex.eprint: https://onlinelibrary.wiley.com/doi/pdf/10.1002/asna.200510446},
	pages = {878--881},
}

@misc{rivilla2026,
	title = {Aromatic rings in the {Central} {Molecular} {Zone}: {Benzonitrile}},
	shorttitle = {Aromatic rings in the {Central} {Molecular} {Zone}},
	url = {http://arxiv.org/abs/2604.24510},
	doi = {10.48550/arXiv.2604.24510},
	urldate = {2026-05-08},
	publisher = {arXiv},
	author = {Rivilla, V. M. and Andrés, D. San and Sanz-Novo, M. and Colzi, L. and Jiménez-Serra, I. and López-Gallifa, A. and Martínez-Henares, A. and Megías, A. and Martín, S. and Tercero, B. and Zeng, S. and Loreau, J. and Khalifa, M. Ben and Requena-Torres, M. A. and Vicente, P. de},
	month = apr,
	year = {2026},
	note = {arXiv:2604.24510 [astro-ph]},
}

@article{agundez2026,
	title = {{TMC}-1: {Probing} the {Onset} of {Chemical} {Complexity} in {Space}},
	shorttitle = {{TMC}-1},
	url = {https://doi.org/10.1021/acsearthspacechem.6c00026},
	doi = {10.1021/acsearthspacechem.6c00026},
	urldate = {2026-05-08},
	journal = {ACS Earth and Space Chemistry},
	publisher = {American Chemical Society},
	author = {Agúndez, Marcelino and Cernicharo, José},
	month = apr,
	year = {2026},
}

@article{tokuda2026,
	title = {{ALMA} {Band} 9 {CO}(6–5) {Reveals} a {Warm} {Ring} {Structure} {Associated} with the {Embedded} {Protostar} in the {Cold} {Dense} {Core} {MC} 27/{L1521F}},
	volume = {1001},
	issn = {2041-8205},
	url = {https://doi.org/10.3847/2041-8213/ae47ec},
	doi = {10.3847/2041-8213/ae47ec},
	language = {en},
	number = {1},
	urldate = {2026-04-26},
	journal = {The Astrophysical Journal Letters},
	publisher = {The American Astronomical Society},
	author = {Tokuda, Kazuki and Omura, Mitsuki and Harada, Naoto and Shoshi, Ayumu and Fukaya, Naofumi and Onishi, Toshikazu and Tachihara, Kengo and Saigo, Kazuya and Matsumoto, Tomoaki and Fukui, Yasuo and Kawamura, Akiko and Machida, Masahiro N.},
	month = apr,
	year = {2026},
	pages = {L1},
}

@article{levey2022,
	title = {{PAH} {Growth} in {Flames} and {Space}: {Formation} of the {Phenalenyl} {Radical}},
	volume = {126},
	issn = {1089-5639},
	shorttitle = {{PAH} {Growth} in {Flames} and {Space}},
	url = {https://doi.org/10.1021/acs.jpca.1c08310},
	doi = {10.1021/acs.jpca.1c08310},
	number = {1},
	urldate = {2026-05-03},
	journal = {The Journal of Physical Chemistry A},
	publisher = {American Chemical Society},
	author = {Levey, Zachariah D. and Laws, Benjamin A. and Sundar, Srivathsan P. and Nauta, Klaas and Kable, Scott H. and da Silva, Gabriel and Stanton, John F. and Schmidt, Timothy W.},
	month = jan,
	year = {2022},
	pages = {101--108},
}

@article{zhao2020,
	title = {Gas phase formation of phenalene via 10π-aromatic, resonantly stabilized free radical intermediates},
	volume = {22},
	issn = {1463-9084},
	url = {https://pubs.rsc.org/en/content/articlelanding/2020/cp/d0cp02216k},
	doi = {10.1039/D0CP02216K},
	language = {en},
	number = {27},
	urldate = {2026-02-11},
	journal = {Physical Chemistry Chemical Physics},
	publisher = {The Royal Society of Chemistry},
	author = {Zhao, Long and Kaiser, Ralf I. and Lu, Wenchao and Ahmed, Musahid and Oleinikov, Artem D. and Azyazov, Valeriy N. and Mebel, Alexander M. and Howlader, A. Hasan and Wnuk, Stanislaw F.},
	month = jul,
	year = {2020},
	pages = {15381--15388},
}

@phdthesis{chitsazzadeh2014,
	type = {Ph.{D}. thesis},
	title = {Internal {Physical} and {Chemical} {Characteristics} of {Starless} {Cores} on the {Brink} of {Gravitational} {Collapse}},
	url = {https://ui.adsabs.harvard.edu/abs/2014PhDT.......423C},
	urldate = {2026-04-26},
	author = {Chitsazzadeh, Shadi},
	month = aug,
	year = {2014},
	note = {ADS Bibcode: 2014PhDT.......423C},
}

@misc{molsim2024,
	title = {molsim},
	url = {https://doi.org/10.5281/zenodo.12697227},
	doi = {10.5281/zenodo.12697227},
	publisher = {Zenodo},
	author = {McGuire, Brett A. and Xue, Ci and Lee, Kin Long Kelvin and El-Abd, Samer and Loomis, Ryan A.},
	month = jul,
	year = {2024},
}

@article{mccarthy2026a,
	title = {Spectroscopic {Studies} of {Polycyclic} {Aromatic} {Hydrocarbons}: {Interstellar} {Aromatic} {Chemistry} {Revealed}},
	shorttitle = {Spectroscopic {Studies} of {Polycyclic} {Aromatic} {Hydrocarbons}},
	url = {https://www.annualreviews.org/content/journals/10.1146/annurev-physchem-082324-010544},
	doi = {10.1146/annurev-physchem-082324-010544},
	language = {en},
	urldate = {2026-02-17},
	publisher = {Annual Reviews},
	author = {McCarthy, Michael C. and McGuire, Brett A.},
	month = feb,
	year = {2026},
}

@article{bourke2006,
	title = {The {Spitzer} c2d {Survey} of {Nearby} {Dense} {Cores}. {II}. {Discovery} of a {Low}-{Luminosity} {Object} in the “{Evolved} {Starless} {Core}” {L1521F}},
	volume = {649},
	issn = {0004-637X},
	url = {https://iopscience.iop.org/article/10.1086/508161/meta},
	doi = {10.1086/508161},
	language = {en},
	number = {1},
	urldate = {2025-07-24},
	journal = {The Astrophysical Journal},
	publisher = {IOP Publishing},
	author = {Bourke, Tyler L. and Myers, Philip C. and Ii, Neal J. Evans and Dunham, Michael M. and Kauffmann, Jens and Shirley, Yancy L. and Crapsi, Antonio and Young, Chadwick H. and Huard, Tracy L. and Brooke, Timothy Y. and Chapman, Nicholas and Cieza, Lucas and Lee, Chang Won and Teuben, Peter and Wahhaj, Zahed},
	month = sep,
	year = {2006},
	pages = {L37},
}

@article{fuentetaja2026,
	title = {Discovery of {1H}-cyclopent[cd]indene (c-{C11H8}) in {TMC}-1 with the {QUIJOTE} line survey: {A} new three-ringed polycyclic aromatic hydrocarbon},
	volume = {706},
	copyright = {© The Authors 2026},
	issn = {0004-6361, 1432-0746},
	shorttitle = {Discovery of {1H}-cyclopent[cd]indene (c-{C11H8}) in {TMC}-1 with the {QUIJOTE} line survey},
	url = {https://www.aanda.org/articles/aa/abs/2026/02/aa58398-25/aa58398-25.html},
	doi = {10.1051/0004-6361/202558398},
	language = {en},
	urldate = {2026-03-27},
	journal = {Astronomy \& Astrophysics},
	publisher = {EDP Sciences},
	author = {Fuentetaja, R. and Cabezas, C. and Agúndez, M. and Tercero, B. and Marcelino, N. and Vicente, P. de and Cernicharo, J.},
	month = feb,
	year = {2026},
	pages = {L10},
}

@article{endres2016,
	series = {New {Visions} of {Spectroscopic} {Databases}, {Volume} {II}},
	title = {The {Cologne} {Database} for {Molecular} {Spectroscopy}, {CDMS}, in the {Virtual} {Atomic} and {Molecular} {Data} {Centre}, {VAMDC}},
	volume = {327},
	issn = {0022-2852},
	url = {https://www.sciencedirect.com/science/article/pii/S0022285216300340},
	doi = {10.1016/j.jms.2016.03.005},
	urldate = {2024-01-29},
	journal = {Journal of Molecular Spectroscopy},
	author = {Endres, Christian P. and Schlemmer, Stephan and Schilke, Peter and Stutzki, Jürgen and Müller, Holger S. P.},
	month = sep,
	year = {2016},
	pages = {95--104},
}

@article{sita2022,
	title = {Discovery of {Interstellar} 2-{Cyanoindene} (2-{C9H7CN}) in {GOTHAM} {Observations} of {TMC}-1},
	volume = {938},
	issn = {2041-8205},
	url = {https://dx.doi.org/10.3847/2041-8213/ac92f4},
	doi = {10.3847/2041-8213/ac92f4},
	language = {en},
	number = {2},
	urldate = {2023-01-24},
	journal = {The Astrophysical Journal Letters},
	publisher = {The American Astronomical Society},
	author = {Sita, Madelyn L. and Changala, P. Bryan and Xue, Ci and Burkhardt, Andrew M. and Shingledecker, Christopher N. and Lee, Kin Long Kelvin and Loomis, Ryan A. and Momjian, Emmanuel and Siebert, Mark A. and Gupta, Divita and Herbst, Eric and Remijan, Anthony J. and McCarthy, Michael C. and Cooke, Ilsa R. and McGuire, Brett A.},
	month = oct,
	year = {2022},
	pages = {L12},
}

@article{burkhardt2021,
	title = {Discovery of the {Pure} {Polycyclic} {Aromatic} {Hydrocarbon} {Indene} (c-{C9H8}) with {GOTHAM} {Observations} of {TMC}-1},
	volume = {913},
	issn = {2041-8205},
	url = {https://doi.org/10.3847/2041-8213/abfd3a},
	doi = {10.3847/2041-8213/abfd3a},
	language = {en},
	number = {2},
	urldate = {2022-09-19},
	journal = {The Astrophysical Journal Letters},
	publisher = {American Astronomical Society},
	author = {Burkhardt, Andrew M. and Lee, Kin Long Kelvin and Changala, P. Bryan and Shingledecker, Christopher N. and Cooke, Ilsa R. and Loomis, Ryan A. and Wei, Hongji and Charnley, Steven B. and Herbst, Eric and McCarthy, Michael C. and McGuire, Brett A.},
	month = may,
	year = {2021},
	note = {tex.ids= Burkhardt2021},
	pages = {L18},
}

@article{cabezas2025,
	title = {Discovery of interstellar phenalene (c-{C13H10}): {A} new piece in the chemical puzzle of {PAHs} in space},
	volume = {701},
	copyright = {© The Authors 2025},
	issn = {0004-6361, 1432-0746},
	shorttitle = {Discovery of interstellar phenalene (c-{C13H10})},
	url = {https://www.aanda.org/articles/aa/abs/2025/09/aa56687-25/aa56687-25.html},
	doi = {10.1051/0004-6361/202556687},
	language = {en},
	urldate = {2025-10-09},
	journal = {Astronomy \& Astrophysics},
	publisher = {EDP Sciences},
	author = {Cabezas, C. and Agúndez, M. and Pérez, C. and Villar-Castro, D. and Molpeceres, G. and Pérez, D. and Steber, A. L. and Fuentetaja, R. and Tercero, B. and Marcelino, N. and Lesarri, A. and Vicente, P. de and Cernicharo, J.},
	month = sep,
	year = {2025},
	pages = {L8},
}

@article{cernicharo2026,
	title = {Discovery of two new isomers of cyanoacenaphthylene ({C12H7CN}) in the {Taurus} molecular cloud 1 with the {QUIJOTE} line survey},
	volume = {705},
	copyright = {© The Authors 2026},
	issn = {0004-6361, 1432-0746},
	url = {https://www.aanda.org/articles/aa/abs/2026/01/aa57893-25/aa57893-25.html},
	doi = {10.1051/0004-6361/202557893},
	language = {en},
	urldate = {2026-02-05},
	journal = {Astronomy \& Astrophysics},
	publisher = {EDP Sciences},
	author = {Cernicharo, J. and Tercero, B. and Marcelino, N. and López-Pérez, J. A. and Gallego, J. D. and Tercero, F. and Esplugues, G. and Cabezas, C. and Agúndez, M. and Limeres, C. and Steber, A. L. and Pérez, D. and Pérez, C. and Lesarri, A. and Vicente, P. de},
	month = jan,
	year = {2026},
	pages = {L7},
}

@article{cernicharo2024,
	title = {Discovery of two cyano derivatives of acenaphthylene ({C12H8}) in {TMC}-1 with the {QUIJOTE} line survey},
	volume = {690},
	copyright = {© The Authors 2024},
	issn = {0004-6361, 1432-0746},
	url = {https://www.aanda.org/articles/aa/abs/2024/10/aa52196-24/aa52196-24.html},
	doi = {10.1051/0004-6361/202452196},
	language = {en},
	urldate = {2024-11-19},
	journal = {Astronomy \& Astrophysics},
	publisher = {EDP Sciences},
	author = {Cernicharo, J. and Cabezas, C. and Fuentetaja, R. and Agúndez, M. and Tercero, B. and Janeiro, J. and Juanes, M. and Kaiser, R. I. and Endo, Y. and Steber, A. L. and Pérez, D. and Pérez, C. and Lesarri, A. and Marcelino, N. and Vicente, P. de},
	month = oct,
	year = {2024},
	pages = {L13},
}

@article{cernicharo2021,
	title = {Pure hydrocarbon cycles in {TMC}-1: {Discovery} of ethynyl cyclopropenylidene, cyclopentadiene and indene.},
	volume = {649},
	issn = {0004-6361},
	shorttitle = {Pure hydrocarbon cycles in {TMC}-1},
	url = {https://www.ncbi.nlm.nih.gov/pmc/articles/PMC7611194/},
	doi = {10.1051/0004-6361/202141156},
	urldate = {2023-01-25},
	journal = {Astronomy and astrophysics},
	publisher = {EDP Sciences},
	author = {Cernicharo, J. and Agúndez, M. and Cabezas, C. and Tercero, B. and Marcelino, N. and Pardo, J. R. and de Vicente, P.},
	month = may,
	year = {2021},
	note = {tex.ids= Cernicharo2021},
	pages = {L15},
}

@article{mcguire2021,
	chapter = {Report},
	title = {Detection of two interstellar polycyclic aromatic hydrocarbons via spectral matched filtering},
	volume = {371},
	copyright = {Copyright © 2021 The Authors, some rights reserved; exclusive licensee American Association for the Advancement of Science. No claim to original U.S. Government Works. https://www.sciencemag.org/about/science-licenses-journal-article-reuseThis is an article distributed under the terms of the Science Journals Default License.},
	issn = {0036-8075, 1095-9203},
	url = {https://science.sciencemag.org/content/371/6535/1265},
	doi = {10.1126/science.abb7535},
	language = {en},
	number = {6535},
	urldate = {2021-03-25},
	journal = {Science},
	publisher = {American Association for the Advancement of Science},
	author = {McGuire, Brett A. and Loomis, Ryan A. and Burkhardt, Andrew M. and Lee, Kin Long Kelvin and Shingledecker, Christopher N. and Charnley, Steven B. and Cooke, Ilsa R. and Cordiner, Martin A. and Herbst, Eric and Kalenskii, Sergei and Siebert, Mark A. and Willis, Eric R. and Xue, Ci and Remijan, Anthony J. and McCarthy, Michael C.},
	month = mar,
	year = {2021},
	pages = {1265--1269},
}

@article{mcguire2018a,
	title = {{\textless}pre{\textgreater}{Detection} of the aromatic molecule benzonitrile (c-{C}\$\_6\${H}\$\_5\${CN}) in the interstellar medium{\textless}/pre{\textgreater}},
	volume = {359},
	copyright = {Copyright © 2018 The Authors, some rights reserved; exclusive licensee American Association for the Advancement of Science. No claim to original U.S. Government Works. http://www.sciencemag.org/about/science-licenses-journal-article-reuseThis is an article distributed under the terms of the Science Journals Default License.},
	issn = {0036-8075, 1095-9203},
	url = {http://science.sciencemag.org/content/359/6372/202},
	doi = {10.1126/science.aao4890},
	language = {en},
	number = {6372},
	urldate = {2018-04-03},
	journal = {Science},
	author = {McGuire, Brett A. and Burkhardt, Andrew M. and Kalenskii, Sergei and Shingledecker, Christopher N. and Remijan, Anthony J. and Herbst, Eric and McCarthy, Michael C.},
	month = jan,
	year = {2018},
	pages = {202--205},
}

@article{wenzel2025c,
	title = {Discovery of the {Seven}-ring {Polycyclic} {Aromatic} {Hydrocarbon} {Cyanocoronene} ({C24H11CN}) in {GOTHAM} {Observations} of {TMC}-1},
	volume = {984},
	issn = {2041-8205},
	url = {https://dx.doi.org/10.3847/2041-8213/adc911},
	doi = {10.3847/2041-8213/adc911},
	language = {en},
	number = {1},
	urldate = {2025-04-30},
	journal = {The Astrophysical Journal Letters},
	publisher = {The American Astronomical Society},
	author = {Wenzel, Gabi and Gong, Siyuan and Xue, Ci and Changala, P. Bryan and Holdren, Martin S. and Speak, Thomas H. and Stewart, D. Archie and Fried, Zachary T. P. and Willis, Reace H. J. and Bergin, Edwin A. and Burkhardt, Andrew M. and Byrne, Alex N. and Charnley, Steven B. and Lipnicky, Andrew and Loomis, Ryan A. and Shingledecker, Christopher N. and Cooke, Ilsa R. and McCarthy, Michael C. and Remijan, Anthony J. and Wendlandt, Alison E. and McGuire, Brett A.},
	month = apr,
	year = {2025},
	pages = {L36},
}

@article{wenzel2025,
	title = {Detections of interstellar aromatic nitriles 2-cyanopyrene and 4-cyanopyrene in {TMC}-1},
	volume = {9},
	copyright = {2024 The Author(s)},
	issn = {2397-3366},
	url = {https://www.nature.com/articles/s41550-024-02410-9},
	doi = {10.1038/s41550-024-02410-9},
	language = {en},
	number = {2},
	urldate = {2025-02-19},
	journal = {Nature Astronomy},
	author = {Wenzel, Gabi and Speak, Thomas H. and Changala, P. Bryan and Willis, Reace H. J. and Burkhardt, Andrew M. and Zhang, Shuo and Bergin, Edwin A. and Byrne, Alex N. and Charnley, Steven B. and Fried, Zachary T. P. and Gupta, Harshal and Herbst, Eric and Holdren, Martin S. and Lipnicky, Andrew and Loomis, Ryan A. and Shingledecker, Christopher N. and Xue, Ci and Remijan, Anthony J. and Wendlandt, Alison E. and McCarthy, Michael C. and Cooke, Ilsa R. and McGuire, Brett A.},
	month = feb,
	year = {2025},
	pages = {262--270},
}

@article{wenzel2024,
	title = {Detection of interstellar 1-cyanopyrene: {A} four-ring polycyclic aromatic hydrocarbon},
	volume = {386},
	shorttitle = {Detection of interstellar 1-cyanopyrene},
	url = {https://www.science.org/doi/10.1126/science.adq6391},
	doi = {10.1126/science.adq6391},
	number = {6723},
	urldate = {2024-11-21},
	journal = {Science},
	publisher = {American Association for the Advancement of Science},
	author = {Wenzel, Gabi and Cooke, Ilsa R. and Changala, P. Bryan and Bergin, Edwin A. and Zhang, Shuo and Burkhardt, Andrew M. and Byrne, Alex N. and Charnley, Steven B. and Cordiner, Martin A. and Duffy, Miya and Fried, Zachary T. P. and Gupta, Harshal and Holdren, Martin S. and Lipnicky, Andrew and Loomis, Ryan A. and Shay, Hannah Toru and Shingledecker, Christopher N. and Siebert, Mark A. and Stewart, D. Archie and Willis, Reace H. J. and Xue, Ci and Remijan, Anthony J. and Wendlandt, Alison E. and McCarthy, Michael C. and McGuire, Brett A.},
	month = nov,
	year = {2024},
	pages = {810--813},
}

@article{subramani2025,
	title = {Recurrent {Fluorescence} of {Polycyclic} {Aromatic} {Hydrocarbon} {Isomers}: {A} {Comparative} {Study}},
	shorttitle = {Recurrent {Fluorescence} of {Polycyclic} {Aromatic} {Hydrocarbon} {Isomers}},
	url = {https://doi.org/10.1021/acsearthspacechem.5c00283},
	doi = {10.1021/acsearthspacechem.5c00283},
	urldate = {2025-11-25},
	journal = {ACS Earth and Space Chemistry},
	publisher = {American Chemical Society},
	author = {Subramani, Arun and Bull, James N. and Cederquist, Henrik and Martini, Paul and Schmidt, Henning T. and Zettergren, Henning and Stockett, Mark H.},
	month = nov,
	year = {2025},
}

@article{xue2025,
	title = {The {Molecular} {Inventory} of {TMC}-1 with {GOTHAM} {Observations}},
	volume = {281},
	issn = {0067-0049},
	url = {https://doi.org/10.3847/1538-4365/ae04e5},
	doi = {10.3847/1538-4365/ae04e5},
	language = {en},
	number = {1},
	urldate = {2025-10-23},
	journal = {The Astrophysical Journal Supplement Series},
	publisher = {The American Astronomical Society},
	author = {Xue, Ci and Byrne, Alex N. and Morgan, Larry and Wenzel, Gabi and Changala, P. Bryan and Fried, Zachary T. P. and Loomis, Ryan A. and Remijan, Anthony and Bergin, Edwin A. and Cooke, Ilsa R. and Frayer, David and Burkhardt, Andrew M. and Charnley, Steven B. and Cordiner, Martin A. and Lipnicky, Andrew and McCarthy, Michael C. and McGuire, Brett A.},
	month = oct,
	year = {2025},
	pages = {9},
}

@article{cernicharo2021b,
	title = {Discovery of benzyne, o-{C6H4}, in {TMC}-1 with the {QUIJOTE} line survey},
	volume = {652},
	copyright = {© ESO 2021},
	issn = {0004-6361, 1432-0746},
	url = {https://www.aanda.org/articles/aa/abs/2021/08/aa41660-21/aa41660-21.html},
	doi = {10.1051/0004-6361/202141660},
	language = {en},
	urldate = {2024-05-09},
	journal = {Astronomy \& Astrophysics},
	publisher = {EDP Sciences},
	author = {Cernicharo, J. and Agúndez, M. and Kaiser, R. I. and Cabezas, C. and Tercero, B. and Marcelino, N. and Pardo, J. R. and Vicente, P. de},
	month = aug,
	year = {2021},
	pages = {L9},
}

@article{cooke2020,
	title = {Benzonitrile as a {Proxy} for {Benzene} in the {Cold} {ISM}: {Low}-temperature {Rate} {Coefficients} for {CN} + {C6H6}},
	volume = {891},
	issn = {2041-8205},
	shorttitle = {Benzonitrile as a {Proxy} for {Benzene} in the {Cold} {ISM}},
	url = {https://dx.doi.org/10.3847/2041-8213/ab7a9c},
	doi = {10.3847/2041-8213/ab7a9c},
	language = {en},
	number = {2},
	urldate = {2024-05-06},
	journal = {The Astrophysical Journal Letters},
	publisher = {The American Astronomical Society},
	author = {Cooke, Ilsa R. and Gupta, Divita and Messinger, Joseph P. and Sims, Ian R.},
	month = mar,
	year = {2020},
	pages = {L41},
}

@article{mcguire2020,
	title = {Early {Science} from {GOTHAM}: {Project} {Overview}, {Methods}, and the {Detection} of {Interstellar} {Propargyl} {Cyanide} ({HCCCH2CN}) in {TMC}-1},
	volume = {900},
	issn = {2041-8205},
	shorttitle = {Early {Science} from {GOTHAM}},
	url = {https://dx.doi.org/10.3847/2041-8213/aba632},
	doi = {10.3847/2041-8213/aba632},
	language = {en},
	number = {1},
	urldate = {2024-05-05},
	journal = {The Astrophysical Journal Letters},
	publisher = {The American Astronomical Society},
	author = {McGuire, Brett A. and Burkhardt, Andrew M. and Loomis, Ryan A. and Shingledecker, Christopher N. and Lee, Kin Long Kelvin and Charnley, Steven B. and Cordiner, Martin A. and Herbst, Eric and Kalenskii, Sergei and Momjian, Emmanuel and Willis, Eric R. and Xue, Ci and Remijan, Anthony J. and McCarthy, Michael C.},
	month = sep,
	year = {2020},
	pages = {L10},
}

@article{frenklach2024,
	title = {Phenalenyl growth reactions and implications for prenucleation chemistry of aromatics in flames},
	issn = {1463-9084},
	url = {https://pubs.rsc.org/en/content/articlelanding/2024/cp/d4cp00096j},
	doi = {10.1039/D4CP00096J},
	language = {en},
	urldate = {2024-04-29},
	journal = {Physical Chemistry Chemical Physics},
	publisher = {The Royal Society of Chemistry},
	author = {Frenklach, Michael and Jasper, Ahren W. and Mebel, Alexander M.},
	month = mar,
	year = {2024},
}

@article{reizer2022,
	title = {Formation and growth mechanisms of polycyclic aromatic hydrocarbons: {A} mini-review},
	volume = {291},
	issn = {0045-6535},
	shorttitle = {Formation and growth mechanisms of polycyclic aromatic hydrocarbons},
	url = {https://www.sciencedirect.com/science/article/pii/S0045653521032653},
	doi = {10.1016/j.chemosphere.2021.132793},
	urldate = {2024-04-04},
	journal = {Chemosphere},
	author = {Reizer, Edina and Viskolcz, Béla and Fiser, Béla},
	month = mar,
	year = {2022},
	pages = {132793},
}

@article{burkhardt2021a,
	title = {Ubiquitous aromatic carbon chemistry at the earliest stages of star formation},
	volume = {5},
	copyright = {2021 The Author(s), under exclusive licence to Springer Nature Limited},
	issn = {2397-3366},
	url = {https://www.nature.com/articles/s41550-020-01253-4},
	doi = {10.1038/s41550-020-01253-4},
	language = {en},
	number = {2},
	urldate = {2024-04-01},
	journal = {Nature Astronomy},
	publisher = {Nature Publishing Group},
	author = {Burkhardt, Andrew M. and Loomis, Ryan A. and Shingledecker, Christopher N. and Lee, Kin Long Kelvin and Remijan, Anthony J. and McCarthy, Michael C. and McGuire, Brett A.},
	month = feb,
	year = {2021},
	pages = {181--187},
}

@article{agundez2023,
	title = {Aromatic cycles are widespread in cold clouds},
	volume = {677},
	copyright = {© The Authors 2023},
	issn = {0004-6361, 1432-0746},
	url = {https://www.aanda.org/articles/aa/abs/2023/09/aa47524-23/aa47524-23.html},
	doi = {10.1051/0004-6361/202347524},
	language = {en},
	urldate = {2024-04-01},
	journal = {Astronomy \& Astrophysics},
	publisher = {EDP Sciences},
	author = {Agúndez, M. and Marcelino, N. and Tercero, B. and Cernicharo, J.},
	month = sep,
	year = {2023},
	pages = {L13},
}

@article{zhao2018,
	title = {Pyrene synthesis in circumstellar envelopes and its role in the formation of {2D} nanostructures},
	volume = {2},
	copyright = {2018 The Author(s)},
	issn = {2397-3366},
	url = {https://www.nature.com/articles/s41550-018-0399-y},
	doi = {10.1038/s41550-018-0399-y},
	language = {en},
	number = {5},
	urldate = {2024-03-28},
	journal = {Nature Astronomy},
	publisher = {Nature Publishing Group},
	author = {Zhao, Long and Kaiser, Ralf I. and Xu, Bo and Ablikim, Utuq and Ahmed, Musahid and Joshi, Dharati and Veber, Gregory and Fischer, Felix R. and Mebel, Alexander M.},
	month = may,
	year = {2018},
	pages = {413--419},
}

@article{rap2024,
	title = {Ionic {Fragmentation} {Products} of {Benzonitrile} as {Important} {Intermediates} in the {Growth} of {Polycyclic} {Aromatic} {Hydrocarbons}},
	issn = {1463-9084},
	url = {https://pubs.rsc.org/en/content/articlelanding/2024/cp/d3cp05574d},
	doi = {10.1039/D3CP05574D},
	language = {en},
	urldate = {2024-02-12},
	journal = {Physical Chemistry Chemical Physics},
	publisher = {The Royal Society of Chemistry},
	author = {Rap, Daniël Benjamin and Schrauwen, Johanna G. M. and Redlich, Britta and Brünken, Sandra},
	month = feb,
	year = {2024},
}

@article{loomis2021,
	title = {An investigation of spectral line stacking techniques and application to the detection of {HC11N}},
	volume = {5},
	copyright = {2021 The Author(s), under exclusive licence to Springer Nature Limited},
	issn = {2397-3366},
	url = {https://www.nature.com/articles/s41550-020-01261-4},
	doi = {10.1038/s41550-020-01261-4},
	language = {en},
	number = {2},
	urldate = {2024-01-29},
	journal = {Nature Astronomy},
	publisher = {Nature Publishing Group},
	author = {Loomis, Ryan A. and Burkhardt, Andrew M. and Shingledecker, Christopher N. and Charnley, Steven B. and Cordiner, Martin A. and Herbst, Eric and Kalenskii, Sergei and Lee, Kin Long Kelvin and Willis, Eric R. and Xue, Ci and Remijan, Anthony J. and McCarthy, Michael C. and McGuire, Brett A.},
	month = feb,
	year = {2021},
	note = {Number: 2},
	pages = {188--196},
}
\bibliographystyle{aasjournalv7}

\end{document}